\documentclass[11pt]{article}

\usepackage[margin=1in]{geometry}
\usepackage[T1]{fontenc}
\usepackage[utf8]{inputenc}
\usepackage{amsmath,amssymb,bm,amsthm}
\usepackage{booktabs,tabularx,array,multirow}
\usepackage{graphicx}
\usepackage{subcaption}
\usepackage{microtype}
\usepackage{hyperref}
\hypersetup{hidelinks}
\usepackage{url}
\usepackage{float}
\usepackage[authoryear,round]{natbib}
\newcommand{\projectgraphic}[2][]{\includegraphics[#1]{#2}}
\newcommand{\R}{\mathbb{R}}
\newcommand{\E}{\mathbb{E}}
\newcommand{\Var}{\mathrm{Var}}

\newcommand{\trans}{^{\mathsf T}}

\newtheorem{theorem}{Theorem}

\newtheorem{lemma}{Lemma}
\newtheorem{assumption}{Assumption}

\title{Bayesian Semiparametric Regression for Count Data Using Zero Altered Moment Models}
\author{
Sujit K. Ghosh\\
\small Department of Statistics, North Carolina State University\\
\small \texttt{sujit\_ghosh@ncsu.edu}
\and
Timothy Boakye\thanks{Corresponding author: \texttt{tboakye@ncsu.edu}}\\
\small Department of Statistics, North Carolina State University
\and
Kimberly F. Sellers\\
\small Department of Statistics, North Carolina State University\\
\small \texttt{kimberly\_sellers@ncsu.edu}
}
\date{}

\begin{document}
\maketitle

\begin{abstract}
We develop a Bayesian semiparametric regression framework for count data that can accommodate both underdispersion and overdispersion without specifying a full parametric probability model. The model is built from the first two conditional factorial moments of a zero altered distribution. Covariates may enter the conditional mean and the conditional dispersion separately, thereby allowing the dispersion to vary across observations. We turn the moment restrictions into an adjusted exponentially tilted empirical likelihood (AETEL) criterion and combine this criterion with Gaussian priors to form a Bayesian empirical-likelihood pseudo-posterior. The adjustment improves numerical feasibility and a two-block Metropolis delayed acceptance sampler allows for faster runs of the MCMC chains. This reduces unnecessary exact evaluations without changing the AETEL target. Using existing Bayesian exponentially tilted empirical likelihood (ETEL) theory, we state a first-order posterior normal approximation and verify the moment, smoothness, and Jacobian rank conditions needed for the asymptotic inference. Simulations under zero inflated Poisson and Conway Maxwell Poisson data exhibit small bias, accurate uncertainty estimates, and near nominal interval coverage for both overdispersed and underdispersed counts. Applications to various datasets illustrate how the Bayesian analysis can identify covariate dependent dispersion using finite sample posterior inference.
\end{abstract}


\section{Introduction}
\label{sec:intro}

Count data appear in many applications, including studies of disease events, insurance claims, firm activity, student attendance, and biological counts. Poisson regression is a natural starting point because it gives a simple log linear model for the conditional mean \citep{McCullaghNelder1989,CameronTrivedi2013}. Its main restriction is that the conditional variance must equal the conditional mean. Real data often violate this condition and may be either overdispersed or underdispersed \citep{CameronTrivedi2013,Puig2024}.

Several parametric models relax the Poisson variance restriction. Negative binomial regression is widely used for overdispersed counts \citep{Lawless1987,CameronTrivedi2013}, and the Conway Maxwell Poisson distribution can accommodate both directions of dispersion \citep{ShmueliEtAl2005,SellersShmueli2010}. The latter flexibility can come with additional numerical work because its probability mass function contains a normalizing constant \citep{SellersShmueli2010}. Quasi likelihood methods offer another route by focusing on low order moments instead of a complete probability model \citep{Wedderburn1974,Smyth1989}. This idea is useful when the mean and variance are scientifically meaningful but the full count distribution is difficult to defend.

The zero altered class of \citet{GhoshKim2007} gives a broad moment based description of dispersed counts. Starting from an unspecified equidispersed baseline distribution, the construction changes the mass at zero and rescales the positive counts.  A key feature is that inference need not estimate the complete baseline probability mass function. This makes the framework semiparametric.

A frequentist regression formulation of this zero altered framework was developed using generalized method of moments (GMM) by \citet{GhoshBoakyeSellers2026} using only the first two factorial moments.  The present paper develops a semiparametric Bayesian inferential framework. The motivation is not that GMM lacks point estimates or uncertainty intervals. Rather, the two approaches use the same moment information but in different inferential perspectives. Traditional GMM obtains an estimator by optimizing a criterion and then uses its sampling behavior for inference \citep{Hansen1982,NeweySmith2004}. A Bayesian moment analysis instead combines a likelihood-type criterion with a prior to define a distribution over the parameter space and uses Markov chain Monte Carlo (MCMC) sampling to summarize it \citep{ChernozhukovHong2003}. This can be useful when a nonlinear criterion is difficult to summarize
adequately by the local curvature at a single optimizer, and it gives direct
uncertainty summaries for nonlinear functions of the parameters without
requiring a separate asymptotic distribution to be derived for each such
function. More generally, \citet{ChernozhukovHong2003} show how MCMC methods can be used as an alternative computational route for estimators
defined by general nonlikelihood criteria, including GMM and related extremum
problems.
This does not imply that MCMC methods are uniformly preferable to optimization;
rather, they replace one computational problem with another and are useful when
the resulting distributional summary is itself informative.

Instead of placing a Gaussian approximation directly around the quadratic GMM criterion, we construct an exponentially tilted empirical likelihood (ETEL) from the same estimating equations \citep{Schennach2005,Schennach2007}. Empirical likelihood methods provide likelihood-type inference from moment conditions without introducing a parametric sampling density \citep{Owen2001,QinLawless1994}. Bayesian empirical likelihood combines this criterion, constructed from the observed data, with a prior distribution.\citep{Lazar2003}. Because the empirical-likelihood criterion is not a full sampling density for the counts, the resulting Bayesian object is more precisely a pseudo-posterior; we make this distinction explicit in Section~\ref{sec:aetel}. The exponentially tilted construction also has a nonparametric Bayesian interpretation for moment restrictions \citep{Schennach2005}. For moment condition models, \citet{ChibShinSimoni2018} developed a Bayesian ETEL framework and established large-sample posterior inference theory. In the present regression setting, the Bayesian formulation therefore complements the frequentist GMM. It uses the same first two factorial moments, but it permits posterior sampling, prior regularization when needed, and direct uncertainty summaries for functions such as the covariate dependent dispersion.

A practical difficulty is that ordinary empirical likelihood and exponential tilting can become numerically unstable when zero is outside, or close to the boundary of, the convex hull of the observed moments. As a result, we add a single artificial moment following the adjusted empirical likelihood approach of \citet{ChenVariyathAbraham2008}. The analogous adjusted exponentially tilted construction has the same first-order asymptotic behavior as ordinary exponential tilting and improves feasibility \citep{ZhuEtAl2009}.

Posterior computation creates a second challenge. Each exact AETEL evaluation requires solving for the tilting vector, so repeated evaluation inside a long Markov chain can dominate the computing time. As a result we use a two-block Metropolis delayed acceptance sampler. Exact adjusted exponentially tilted empirical likelihood (AETEL) evaluations from an initial pilot phase are used to build an inverse-distance-weighted surrogate, which is then frozen. The mean and dispersion coefficients are updated in separate blocks. A proposal is first screened by the surrogate; only a proposal that passes this first stage requires evaluation of the exact AETEL criterion. A second-stage correction then preserves the exact AETEL pseudo-posterior target. This follows the delayed acceptance principle of \citet{ChristenFox2005}, where the approximation enhances computational efficiency, it does not change the target distribution being sampled.

The paper has four main contributions. First, it gives a Bayesian regression formulation for the zero altered moment model with separate mean and dispersion effects. Second, it develops an adjusted exponentially tilted empirical likelihood pseudo-posterior that uses only the first two conditional factorial moments. Third, it gives a stable two-block delayed acceptance scheme that reduces exact AETEL evaluations and preserves the intended AETEL pseudo-posterior. Fourth, it studies the method under both overdispersion and underdispersion and applies it to three data examples where dispersion changes with observed covariates.

The rest of the paper is organized as follows. Section~\ref{sec:model} introduces the zero altered regression moments. Section~\ref{sec:aetel} develops the Bayesian adjusted exponentially tilted empirical likelihood and the delayed acceptance sampler. Section~\ref{sec:theory} connects the construction to existing Bayesian ETEL asymptotics and verifies the required moment regularity for the zero altered regression moments. Section~\ref{sec:simulation} presents the simulation study. Section~\ref{sec:applications} gives three applications. Section~\ref{sec:conclusion} concludes. Supporting derivations, computational settings, simulation generation details, verification of the regularity conditions, and Bayesian diagnostics are collected in the appendices of this paper.

\section{Zero Altered Regression and Moment Restrictions}
\label{sec:model}

\subsection{Zero altered representation}

Following the inferential framework of \citet{GhoshKim2007}, let \(U\sim f_0\), where \(f_0\) is an equidispersed probability mass function on \(\mathbb N=\{0,1,2,\ldots\}\), with \(\mu_0=\E_{f_0}(U)\), \(\Var_{f_0}(U)=\mu_0\), and \(\pi_0=f_0(0)\).
For \(\delta\in(-1,1)\), define \(\delta_+=\max(\delta,0)\), \(\delta_-=\max(-\delta,0)\), and
\begin{equation}
\label{eq:za_pmf}
f_\delta(y)=
\begin{cases}
\delta_+^2+(1-\delta^2)\pi_0, & y=0,\\[3pt]
\omega(\delta,\pi_0)f_0(y), & y\ge 1,
\end{cases}
\qquad
\omega(\delta,\pi_0)
=1-\delta_+^2+\delta_-^2\frac{\pi_0}{1-\pi_0}>0.
\end{equation}
Thus, the mass at zero is changed and all positive probabilities are multiplied by the same factor.

As shown by \citet{GhoshKim2007}, if \(Y\sim f_\delta\), then \(
\mu=\E(Y)=\omega\mu_0.
\) Since the baseline distribution is equidispersed, it follows that
\begin{equation}
\label{eq:qvf}
\Var(Y)=\mu+\phi\mu^2,
\qquad
\phi=\frac{1-\omega}{\omega}\in(-1,\infty).
\end{equation}

Hence \(\omega>1\), or equivalently \(\phi<0\), gives underdispersion;
\(\omega=1\), or equivalently \(\phi=0\), gives equidispersion; and \(\omega<1\), or equivalently \(\phi>0\), gives overdispersion.
When \(\delta=0\), we have \(\omega=1\), and the model returns to the
baseline distribution.

The probability mass function in \eqref{eq:za_pmf} provides the starting point for the model, but the Bayesian procedure below is based only on the first two factorial moment restrictions and does not specify or estimate the complete baseline distribution \(f_0\). Thus, \(f_0\) remains an unspecified infinite-dimensional
nuisance component, which gives the model its semiparametric character. The prior is placed on the finite-dimensional regression parameters rather than on \(f_0\). This is consistent with the original zero altered construction of \citet{GhoshKim2007}.

\subsection{Regression moments}

We let \(Y_i\) denote the count response for observation \(i\) and let
\(x_i\in\R^p\) contain the regressors, including an intercept, for both the
mean and dispersion model. We write
\(\theta=(\beta\trans,\gamma\trans)\trans\), with
\(\beta,\gamma\in\R^p\). We model the conditional mean and dispersion via link functions:
\begin{equation}
\label{eq:links}
\mu_i(\beta)=g(x_i\trans\beta),
\qquad
1+\phi_i(\gamma)=h(x_i\trans\gamma),
\end{equation}
where \(g\) and \(h\) map the real line to the positive real line. Our applications use the log links
\(\mu_i=\exp(x_i\trans\beta)\) and
\(\phi_i=\exp(x_i\trans\gamma)-1\), which automatically enforce
\(\phi_i>-1\) as required by \eqref{eq:qvf}. Other suitable link functions
for \(g\) and \(h\) may also be chosen depending on the application.

The factorial moment identity
\(\E\{Y_i(Y_i-1)\mid x_i\}=\Var(Y_i\mid x_i)+\mu_i^2-\mu_i\)
gives
\(\E\{Y_i(Y_i-1)\mid x_i\}=\mu_i^2(1+\phi_i)\).
The varying dispersion model is therefore characterized by the \(2p\)-dimensional moment vector
\begin{equation}
\label{eq:moment}
m_i(\theta)=
\begin{pmatrix}
x_i\{Y_i-\mu_i(\beta)\}\\[4pt]
x_i\left[Y_i(Y_i-1)-\mu_i(\beta)^2h(x_i\trans\gamma)\right]
\end{pmatrix},
\qquad
\E\{m_i(\theta_0)\}=0.
\end{equation}
where, \(\theta_0\) denotes the true parameter value and is assumed to be the unique solution of the population moment equation
\(\E\{m_i(\theta)\}=0\).

Thus the same covariates are used in both blocks of the varying dispersion model. For the constant dispersion fits, the computation uses the corresponding reduced moment system with an intercept-only dispersion component. The free parameter and moment dimension is then \(d=p+1\), whereas the full varying dispersion model has \(d=2p\). The moment restrictions remain the only data model needed for the Bayesian AETEL construction.

\section{Bayesian Adjusted Exponentially Tilted Empirical Likelihood}
\label{sec:aetel}

The ETEL construction begins with the estimating equations and not a fully
specified sampling distribution. We use ETEL because it converts the moment
restrictions into a likelihood-type criterion through exponential tilting and
maintains the moment information that defines the regression model
\citep{Schennach2005,Schennach2007}. This provides a direct way to combine the
estimating equations with a prior and to obtain posterior draws for the
regression parameters and nonlinear functions of them \citep{Lazar2003}.

Bayesian empirical likelihood also has a broader semiparametric Bayesian
interpretation. \citet{Sueishi2024BEL} provides a theoretical bridge between
Bayesian empirical likelihood and semiparametric Bayesian inference. Using
Bernstein--von Mises theory, the limiting Bayesian empirical likelihood
posterior is shown to agree with a semiparametric Bayesian procedure that
places priors on both a finite-dimensional parameter of interest and an
infinite-dimensional nuisance parameter, and also with a parametric Bayesian
procedure based on a least-favorable model.

Relatedly, \citet{BornnShephardSolgi2019} develops a Bayesian nonparametric
framework that combines a nonparametric distribution for the observations
with moment restrictions defining the parameter of interest. Their work
develops the associated probability theory and computational methods,
including methods for simulating from distributions supported on the
constraint manifold.

Compared with using a quadratic objective alone, the ETEL criterion is built
from observation-specific implied probabilities that satisfy the empirical
moment restrictions. This is particularly useful here because the proposed
method is defined through the first two factorial moments and not through a
complete count likelihood. Ordinary ETEL, however, can become infeasible when
zero is outside, or close to the boundary of, the convex hull of the observed
moment vectors. We therefore use an adjusted ETEL construction that appends one
artificial moment before solving for the exponential tilting weights.

\subsection{Adjusted exponential tilting}

For a fixed \(\theta\), write \(d\) for the dimension of the moment vector and define \(\bar m_n(\theta)=n^{-1}\sum_{i=1}^n m_i(\theta)\). We append one artificial moment,
\begin{equation}
\label{eq:adjust}
m_{n+1}^{a}(\theta)
=
-a_n\bar m_n(\theta),
\qquad
a_n
=
\max\left\{1,\frac{1}{2}\log n\right\}.
\end{equation}
For \(n\geq 8\), which includes all simulation settings and real-data applications considered here, this choice reduces to \(a_n=\frac{1}{2}\log n\). The adjustment is small relative to the sample size but makes the empirical
moment constraint feasible over a much larger part of the parameter space.
This choice follows the adjusted empirical likelihood construction of
\citet{ChenVariyathAbraham2008}. Adjusted exponential tilting preserves the
first-order properties of the original exponential tilting estimator
\citep{ZhuEtAl2009}. The logarithmic choice in \eqref{eq:adjust} is used in the simulation study and in the real data application. Fixed coefficients also satisfy the first-order requirement \(a_n=o(\sqrt n)\).

With \(N=n+1\), write \(m_i^a(\theta)=m_i(\theta)\) for \(i\le n\), with
\(m_N^a(\theta)\) given by \eqref{eq:adjust}. For a tilting vector
\(\lambda\in\R^{d}\), define \citep{Schennach2007}
\begin{equation}
\label{eq:weights}
p_i(\theta,\lambda)
=
\frac{\exp\{\lambda\trans m_i^a(\theta)\}}
{\displaystyle\sum_{j=1}^{N}
\exp\{\lambda\trans m_j^a(\theta)\}},
\qquad
i=1,\ldots,N.
\end{equation}

For each \(\theta\), the tilting vector
\(\lambda=\lambda(\theta)\) is chosen to satisfy
\begin{equation}
\label{eq:tilting}
\bar m_w(\lambda;\theta)
=
\sum_{i=1}^{N}
p_i(\theta,\lambda)m_i^a(\theta)
=
0.
\end{equation}

The AETEL criterion is \(L_A(\theta)=\prod_{i=1}^{N}p_i\{\theta,\lambda(\theta)\}\). For computation, we use its log form
\begin{equation}
\label{eq:log_aetel}
\ell_A(\theta)
=
\sum_{i=1}^{N} b_i(\theta)
-
N\log\left\{
\sum_{j=1}^{N}\exp\bigl(b_j(\theta)\bigr)
\right\},
\qquad
b_i(\theta)
=
\lambda(\theta)\trans m_i^a(\theta).
\end{equation}
To improve numerical stability, the logarithmic sum in \eqref{eq:log_aetel}
is evaluated using the  log-sum-exp (LSE) identity,
\[
\log\left(\sum_{j=1}^{N} e^{b_j}\right)
=
b_{\max}
+
\log\left(\sum_{j=1}^{N} e^{b_j-b_{\max}}\right),
\qquad
b_{\max}=\max_j b_j,
\]
which avoids numerical overflow when some \(b_j\) values are large.

\subsection{Prior and AETEL pseudo-posterior}

We use Gaussian prior blocks \(\beta\sim N_p(0,\Sigma_\beta)\) and \(\gamma\sim N_p(0,\Sigma_\gamma)\), where the covariance matrices are positive definite. In the simulation study, the two parameter blocks use the same weakly regularizing prior specification. For the real data analyses, we use Gaussian regularization on the standardized working scale to improve finite sample numerical stability. In these analyses, the mean block uses diagonal Gaussian precision, and the dispersion block also includes a design scaled precision term. The exact prior forms and application specific settings are given in the computational appendix.

Let \(\mathcal D_n=\{(Y_i,x_i):i=1,\ldots,n\}\) for the observed regression data. Combining the adjusted ETEL criterion with the prior we obtain
\begin{equation}
\label{eq:posterior}
\pi_A(\theta\mid\mathcal D_n)
\propto
L_A(\theta)\,\pi(\theta),
\end{equation}
where \(\theta=(\beta\trans,\gamma\trans)\trans\). Up to an additive constant, the log kernel is \(r_n(\theta)=\ell_A(\theta)-\frac12\beta\trans\Sigma_\beta^{-1}\beta-\frac12\gamma\trans\Sigma_\gamma^{-1}\gamma\).

Because \(L_A(\theta)\) is a likelihood-type empirical criterion constructed from moment restrictions rather than a sampling density for the count response, \(\pi_A(\theta\mid\mathcal D_n)\) is more precisely an AETEL pseudo-posterior, not a posterior arising from a fully specified parametric probability model \citep{Lazar2003,ChibShinSimoni2018}. This distinction is important for the semiparametric formulation: the full count distribution is not specified, and the prior is placed only on the finite-dimensional regression parameters. The large-sample result in Section~\ref{sec:theory} is stated for a fixed prior that is proper, positive, and continuous in a neighborhood of the true parameter value \(\theta_0\).

\subsection{Stable numerical evaluation}

At every exact AETEL posterior evaluation, the tilting equation
\eqref{eq:tilting} must be solved. Define the weighted mean of the adjusted moments by \(\bar m_w(\lambda;\theta)=\sum_{j=1}^{N}p_j(\theta,\lambda)m_j^a(\theta)\). Thus, solving the tilting equation is equivalent to finding \(\lambda\) such that \(\bar m_w(\lambda;\theta)=0\). For \(j=1,\ldots,N\), define the centered adjusted moment \(C_j(\lambda;\theta)=m_j^a(\theta)-\bar m_w(\lambda;\theta)\).
The weighted covariance matrix of the adjusted moments is then
\[
H(\lambda;\theta)
=
\sum_{j=1}^{N}
p_j(\theta,\lambda)
C_j(\lambda;\theta)
C_j(\lambda;\theta)\trans
\in\R^{d\times d}.
\]
This matrix is also the derivative of the weighted moment equation with respect to \(\lambda\), so \(\partial\bar m_w(\lambda;\theta)/\partial\lambda\trans=H(\lambda;\theta)\). We therefore solve \(\bar m_w(\lambda;\theta)=0\) using a Newton update.
Given the current value \(\lambda^{(s)}\), the Newton direction is \(d^{(s)}=H\{\lambda^{(s)};\theta\}^{-1}\bar m_w\{\lambda^{(s)};\theta\}\), and the next value is \(\lambda^{(s+1)}=\lambda^{(s)}-t_s d^{(s)}\), with \(0<t_s\le 1\),
where \(t_s\) is selected by backtracking when a full Newton step does not
reduce the norm of the tilting equation \citep{NocedalWright2006}.

To avoid numerical overflow, the exponential weights in
\eqref{eq:weights} are computed after subtracting their maximum before
exponentiation. To avoid unstable matrix inversion, very small eigenvalues
of \(H(\lambda;\theta)\) are floored before the Newton direction is
computed. The previous solution for \(\lambda\) is used as a warm start
whenever possible. Further numerical details are given in the computational appendix.

\subsection{Two block delayed acceptance sampling}
\label{sec:computation}

The exact adjusted AETEL posterior is more costly to evaluate than a GMM
objective because every exact evaluation requires solving a non-linear equation for the tilting vector. We reduce this cost using a two block Metropolis delayed acceptance
sampler. The mean coefficients \(\beta\) form one block and the
dispersion coefficients \(\gamma\) form the second. During an initial pilot period, proposals are evaluated using the exact
AETEL criterion. The resulting parameter values and exact log criteria
are stored in a cache \(\mathcal C_{N_c}=\{(\theta^{(c)},\ell_A(\theta^{(c)})):c=1,\ldots,N_c\}\), where \(N_c\) is the number of cached exact evaluations. After the pilot
period, the cache is frozen.

For a proposed value \(\theta\), let
\(\mathcal N_k(\theta)\) denote the indices of the \(k\) nearest cached
parameter values. For \(c\in\mathcal N_k(\theta)\), define the inverse distance weight \(w_c(\theta)=\{\|\theta-\theta^{(c)}\|^2+\varepsilon\}^{-\kappa/2}\), where \(\varepsilon>0\) prevents division by zero and \(\kappa>0\) controls how rapidly the weight decreases with distance.
The inverse distance weighted surrogate log criterion is
\[
\widetilde{\ell}_A(\theta)
=
\frac{
\displaystyle
\sum_{c\in\mathcal N_k(\theta)}
w_c(\theta)\ell_A(\theta^{(c)})
}{
\displaystyle
\sum_{c\in\mathcal N_k(\theta)}
w_c(\theta)
}.
\]
Combining this surrogate criterion with the same prior used in the exact AETEL posterior gives \(\widetilde{\pi}(\theta\mid\mathcal D_n)\propto\exp\{\widetilde{\ell}_A(\theta)\}\pi(\theta)\). The surrogate is therefore constructed from nearby exact AETEL
evaluations and is used only as a computational approximation for
screening proposals.

Within each block, a symmetric Gaussian random walk generates a proposal
\(\theta'\). The proposal first passes through the surrogate screen with
probability
\begin{equation}
\label{eq:stage1}
\alpha_1(\theta,\theta')
=
\min\left\{
1,
\frac{\widetilde{\pi}(\theta'\mid\mathcal D_n)}
{\widetilde{\pi}(\theta\mid\mathcal D_n)}
\right\}.
\end{equation}
Only proposals that pass this first stage require a new exact AETEL
evaluation. Conditional on passing the first stage, the second stage
accepts the proposal with probability
\begin{equation}
\label{eq:correction}
\alpha_2(\theta,\theta')
=
\min\left\{
1,
\frac{\pi_A(\theta'\mid\mathcal D_n)}
{\pi_A(\theta\mid\mathcal D_n)}
\frac{\widetilde{\pi}(\theta\mid\mathcal D_n)}
{\widetilde{\pi}(\theta'\mid\mathcal D_n)}
\right\}.
\end{equation}
If the exact tilting solve fails at the proposed state, the proposal is
rejected. For a fixed surrogate, the two-stages in
\eqref{eq:stage1} and \eqref{eq:correction} give the standard delayed
acceptance correction and leave the exact AETEL pseudo-posterior in
\eqref{eq:posterior} invariant \citep{ChristenFox2005}. Thus, the
surrogate enhances computational efficiency but does not replace the
exact AETEL pseudo-posterior target.

The simulation study and the three real data applications use this same
delayed acceptance framework. Their run lengths and proposal tuning
differ because the parameter dimensions and study designs differ, but the AETEL pseudo-posterior target and the two-stage correction remain the same.

Posterior summaries are computed from the retained MCMC output. Suppose
\(\theta^{(1)},\ldots,\theta^{(M)}\) denote the \(M\) posterior draws kept
after burn in and thinning. For the real-data applications, the point summary
reported for each coefficient is the componentwise posterior median. For
parameter \(\theta_j\), this is estimated by
\[
\widehat{\theta}_{j,0.5}
=
\operatorname{median}\left\{
\theta_j^{(1)},\ldots,\theta_j^{(M)}
\right\}.
\]
The repeated Monte Carlo study retains the posterior mean from each replication
for its bias and empirical standard deviation calculations, as used in the
reported simulation results. A \(100(1-\alpha)\%\) highest posterior density
region may be written as
\[
S_\alpha
=
\left\{
\theta\in\Theta:
L_A(\theta)\pi(\theta)\ge c_\alpha
\right\},
\]
where \(c_\alpha>0\) is chosen so that
\[
\frac{
\displaystyle\int_{S_\alpha}
L_A(\theta)\pi(\theta)\,d\theta
}{
\displaystyle\int_{\Theta}
L_A(\theta)\pi(\theta)\,d\theta
}
=
1-\alpha,
\qquad
0<\alpha<1.
\]
For the coefficient summaries, the marginal \(95\%\) Highest Posterior Density (HPD)
intervals are approximated from the retained MCMC draws by taking the shortest interval containing 95 percent of the sampled marginal values for each parameter.

\section{Large Sample Posterior Consistency}
\label{sec:theory}

The large-sample argument uses existing Bayesian ETEL theory. The MCMC analysis itself does not require an asymptotic normal approximation to construct the reported Bayesian intervals or posterior probabilities. The purpose of this section is instead to connect the AETEL construction to the established large-sample theory and to the frequentist GMM analysis, and to verify the model-specific conditions for the zero-altered regression moments in \eqref{eq:moment}. The general Bayesian ETEL inferential framework is based on influential papers by \citet{ChibShinSimoni2018}; our contribution here is to verify the moment envelopes, differentiability, and Jacobian rank for the regression moments used in this paper. Our estimating equations already form a fixed finite-dimensional vector of unconditional moments after multiplication by the regressors.

For the full varying dispersion model, let \(d=2p\), \(\mathcal U_i=(Y_i,x_i\trans)\trans\), \(G_0=\E\{\partial m_i(\theta_0)/\partial\theta\trans\}\), \(S_0=\E\{m_i(\theta_0)m_i(\theta_0)\trans\}\), and \(J_0=G_0\trans S_0^{-1}G_0\). The constant dispersion analyses use the corresponding reduced system with \(d=p+1\).

\begin{assumption}[Regression regularity]
\label{ass:zamregular}
The pairs \(\{\mathcal U_i\}_{i=1}^n\) are independent and identically distributed. The true value \(\theta_0\) is an interior point of a compact parameter set, \(\|x_i\|\le K_x<\infty\) almost surely, and \(\E(Y_i^4)<\infty\). The link functions are twice continuously differentiable and positive on the range of the linear predictors generated by this parameter set.
\end{assumption}

\begin{assumption}[Identification and information]
\label{ass:identinfo}
The population equation \(\E\{m_i(\theta)\}=0\) uniquely identifies \(\theta_0\), and \(S_0\) is positive definite. For the log-link varying dispersion model, define \(A_\beta=\E(\mu_i x_ix_i\trans)\) and \(A_\gamma=\E\{\mu_i^2(1+\phi_i)x_ix_i\trans\}\), evaluated at \(\theta_0\), and assume \(A_\beta\) and \(A_\gamma\) are positive definite. The theoretical prior density \(\pi(\theta)\) is fixed, proper, continuous,
and strictly positive on the parameter space \(\Theta\).
\end{assumption}

\begin{lemma}[Regularity and Jacobian structure of the zero altered moments]
\label{prop:zamregularity}
Under Assumptions~\ref{ass:zamregular}--\ref{ass:identinfo}, the moment vector in \eqref{eq:moment} has a square integrable envelope in a neighborhood of \(\theta_0\), its first derivative has an integrable envelope, and the population Jacobian for the log-link varying dispersion model is
\begin{equation}
\label{eq:G0block}
G_0
=
-\begin{pmatrix}
A_\beta & 0\\
2A_\gamma & A_\gamma
\end{pmatrix}.
\end{equation}
Consequently \(G_0\) has full column rank. The corresponding reduced constant dispersion system has the same block lower triangular form, with a scalar positive dispersion block.
\end{lemma}

The envelope result follows because bounded covariates and compactness bound the linear predictors and the links. The first moment block is bounded by a constant times \(1+|Y_i|\), and the second is bounded by a constant times \(1+Y_i^2\). The finite fourth moment then gives square integrability. For the log links, direct differentiation of the two blocks gives \eqref{eq:G0block}; positive definiteness of its diagonal blocks gives full rank. The appendix gives the detailed bounds and derivative calculation. Identification and positive definiteness of \(S_0\) remain model conditions and are not implied by the envelope argument.

\begin{theorem}[First-order AETEL posterior normal approximation]
\label{thm:bvm}
Suppose Assumptions~\ref{ass:zamregular}--\ref{ass:identinfo} hold, the standard posterior concentration and local ETEL expansion conditions in \citet{ChibShinSimoni2018} hold for this fixed dimensional moment system, and \(a_n=o(\sqrt n)\). Let \(\widehat\theta_A\) denote the local maximizer of the adjusted ETEL criterion. Then the first-order Bayesian ETEL limit, together with the first-order negligibility of the adjustment, gives
\begin{equation}
\label{eq:bvm}
\mathcal L\left\{\sqrt n(\theta-\widehat\theta_A)\mid\mathcal D_n\right\}
\Longrightarrow
N_d(0,J_0^{-1})
\end{equation}
in probability.
\end{theorem}

Theorem~\ref{thm:bvm} is an application of the existing Bayesian ETEL theory of \citet{ChibShinSimoni2018}. The model specific work here is the verification of the moment envelopes, derivative bounds, and Jacobian rank for the zero altered factorial moment regression actually used in the simulations and applications. The artificial moment in \eqref{eq:adjust} is first-order negligible when \(a_n=o(\sqrt n)\), which includes the logarithmic adjustment \citep{ZhuEtAl2009}. The design scaled Gaussian regularizers used in the real data fits are finite sample computational choices.

The result also explains why Bayesian AETEL and frequentist GMM can be close in moderate and large-samples. Both are influenced by the same moment restrictions and the same first-order information \citep{Hansen1982,NeweySmith2004}. The asymptotic normal approximation in \eqref{eq:bvm} is not used to
construct the Bayesian intervals or posterior sign probabilities. The finite samples based on HPD intervals are computed
directly from the retained MCMC posterior draws.

\section{Simulation Study}
\label{sec:simulation}

\subsection{Design}

We compare frequentist GMM with Bayesian AETEL in two settings. Both use one covariate \(x\sim\mathrm{Unif}(-1,1)\) and the log links \(\mu(x)=\exp(\beta_0+\beta_1x)\) and \(1+\phi(x)=\exp(\gamma_0+\gamma_1x)\).
The first setting is overdispersed. We take \(n=200\), \(200\) Monte Carlo replications, \((\beta_0,\beta_1,\gamma_0,\gamma_1)=(\log 3,0.40,0.20,0.20)\), and generate a zero inflated Poisson (ZIP) variable whose first two factorial moments match the target \(\mu(x)\) and \(\phi(x)\). The second setting is underdispersed. We take \(n=200\), \(200\) replications, and \((\beta_0,\beta_1,\gamma_0,\gamma_1)=(0.60,0.10,-0.20,-0.10)\).
A Conway Maxwell Poisson (CMP) distribution is calibrated numerically at each covariate value so that its first two factorial moments equal the same target moments. The choice \(n=200\) examines finite-sample performance at a sample size closer to those in the three applications. The number of replications was chosen to balance Monte Carlo precision with the substantial computational cost of the Bayesian AETEL fits.

The GMM estimator and the Bayesian method use the same moment system based on the first two factorial moment restrictions. Bayesian AETEL uses the same weakly regularizing prior specification for all regression coefficients. For the repeated Monte Carlo runs, posterior sampling uses the two block delayed acceptance scheme described in Section~\ref{sec:aetel}. Each chain uses an exact pilot phase to build the frozen surrogate, followed by the two-stage updates for the mean and dispersion blocks. For Bayesian AETEL, the bias and empirical standard deviation calculations use the posterior mean from each Monte Carlo fit, matching the reported simulation results. We summarize bias, empirical standard deviation (SD), average standard error (SE) or posterior SD, and 95 percent interval coverage. For GMM, coverage is based on Wald confidence intervals, whereas for Bayesian AETEL it is based on 95 \%  highest posterior density (HPD)
credible intervals.

\subsection{Monte Carlo results}

Table~\ref{tab:sim_zip} shows the overdispersed results. Both methods have small bias. For Bayesian AETEL, the posterior mean from each Monte Carlo replicate is used as the point estimate for computing bias and empirical standard deviation. The empirical standard deviations and uncertainty estimates are close, and coverage is near the nominal 0.95 level. The coverage values for the two methods are similar in this experiment.

\begin{table}[H]
\centering
\caption{Overdispersed zero inflated Poisson simulation, \(n=200\), 200 replications.}
\label{tab:sim_zip}
\small
\setlength{\tabcolsep}{4pt}
\begin{tabular}{llrrrrr}
\toprule
Method & Parameter & Truth & Bias & Empirical SD & Avg. SE/SD & Coverage \\
\midrule
GMM & $\beta_0$  & 1.0986 & -0.005674 & 0.05312 & 0.05381 & 0.945 \\
    & $\beta_1$  & 0.4000 & -0.009433 & 0.10310 & 0.09774 & 0.935 \\
    & $\gamma_0$ & 0.2000 & -0.001255 & 0.04542 & 0.04822 & 0.955 \\
    & $\gamma_1$ & 0.2000 &  0.013004 & 0.09695 & 0.08836 & 0.925 \\
\addlinespace
AETEL & $\beta_0$  & 1.0986 & -0.014805 & 0.05334 & 0.05493 & 0.950 \\
      & $\beta_1$  & 0.4000 & -0.008491 & 0.10382 & 0.09851 & 0.940 \\
      & $\gamma_0$ & 0.2000 &  0.008228 & 0.04587 & 0.04929 & 0.950 \\
      & $\gamma_1$ & 0.2000 &  0.020399 & 0.09848 & 0.08947 & 0.925 \\
\bottomrule
\end{tabular}
\end{table}

The results in Table~\ref{tab:sim_cmp} for underdispersed model lead to the same overall conclusion. Bias remains small, and the uncertainty estimates track the Monte Carlo variability. Coverage ranges from 0.920 to 0.985. This case is important because the data are generated from a fully parametric count distribution that is not used in fitting. The Bayesian procedure still recovers the regression parameters because the first two factorial moments are correctly specified.

\begin{table}[H]
\centering
\caption{Underdispersed Conway Maxwell Poisson simulation, \(n=200\), 200 replications.}
\label{tab:sim_cmp}
\small
\setlength{\tabcolsep}{4pt}
\begin{tabular}{llrrrrr}
\toprule
Method & Parameter & Truth & Bias & Empirical SD & Avg. SE/SD & Coverage \\
\midrule
GMM & $\beta_0$  & 0.6000 & -0.009770 & 0.03999 & 0.04310 & 0.970 \\
    & $\beta_1$  & 0.1000 & -0.004637 & 0.07730 & 0.07427 & 0.945 \\
    & $\gamma_0$ & -0.2000 & -0.009317 & 0.04472 & 0.04487 & 0.920 \\
    & $\gamma_1$ & -0.1000 &  0.003881 & 0.08034 & 0.07639 & 0.945 \\
\addlinespace
AETEL & $\beta_0$  & 0.6000 & -0.016480 & 0.04040 & 0.04412 & 0.985 \\
      & $\beta_1$  & 0.1000 & -0.004310 & 0.07741 & 0.07566 & 0.945 \\
      & $\gamma_0$ & -0.2000 &  0.000076 & 0.04587 & 0.04509 & 0.945 \\
      & $\gamma_1$ & -0.1000 &  0.000430 & 0.08395 & 0.07703 & 0.945 \\
\bottomrule
\end{tabular}
\end{table}

These simulations show that the Bayesian procedure works in both directions of dispersion with the same parameterization. Also, the method is not tied to the data generating probability mass function. The zero inflated Poisson and Conway Maxwell Poisson mechanisms are quite different, but both are handled through the same set of moment restrictions.

The average lengths of the 95\% intervals were very similar for FGMM and BAETEL. In the ZIP setting, the average lengths for \((\beta_0,\beta_1,\gamma_0,\gamma_1)\) were \((0.211,0.383,0.189,0.346)\) for FGMM and \((0.215,0.385,0.193,0.350)\) for BAETEL. In the CMP setting, the corresponding average lengths were \((0.169,0.291,0.176,0.299)\) for FGMM and \((0.173,0.297,0.176,0.301)\) for BAETEL. Thus, the two methods produced intervals of comparable width in both settings.

\subsection{Computational behavior}

The delayed acceptance diagnostics show that the surrogate removed a meaningful share of proposals before the exact AETEL solve. The first stage pass rate is the proportion of proposed block updates that pass the surrogate screen and therefore require an exact AETEL evaluation, and the final acceptance rate is the proportion of proposals accepted after both stages.

In the ZIP experiment, the average first stage pass rates across replications were 0.4087 for the mean block and 0.4850 for the dispersion block. Thus, the surrogate avoided an exact AETEL evaluation for approximately 59.1 percent and 51.5 percent of the proposed mean and dispersion updates, respectively. The corresponding final acceptance rates were 0.3551 and 0.4194.

In the CMP experiment, the first stage pass rates were 0.3850 and 0.4996 for the mean and dispersion blocks, respectively, while the final acceptance rates were 0.3392 and 0.4422. Hence, approximately 61.5 percent of the mean block proposals and 50.0 percent of the dispersion block proposals were rejected without requiring an exact AETEL evaluation.

These results illustrate the computational role of delayed acceptance. Since an exact AETEL evaluation requires solving the tilting equations, screening proposals with the less expensive surrogate reduces the number of exact AETEL evaluations required during MCMC. The second stage correction is then applied to proposals that pass the surrogate screen so that the sampler continues to target the exact AETEL pseudo-posterior.

\section{Applications}
\label{sec:applications}

We analyze three data sets that were also used to study the corresponding
frequentist moment model. In every case, the conditional mean uses a log
link. We fit both a constant dispersion model and a varying dispersion
model. Continuous predictors are centered or standardized when this
improves computation.  All application fits use the delayed acceptance sampler described in
Section~\ref{sec:computation}. The computational advantage comes from using
the inexpensive surrogate to screen proposals before evaluating the exact
AETEL pseudo-posterior. Thus, an exact tilting solve is required only for proposals
that pass the first stage, which reduce the number of computationally expensive
exact AETEL evaluations while the second stage correction preserves the
exact AETEL pseudo-posterior target.

For each coefficient, we show the posterior probability
\(\Pr(\theta_j>0\mid\mathcal D_n)\), posterior standard deviation (SD),
posterior median, and 95 percent highest posterior density (HPD) interval.
Only \(\Pr(\theta_j>0\mid\mathcal D_n)\) is reported because
\(\Pr(\theta_j<0\mid\mathcal D_n)=1-\Pr(\theta_j>0\mid\mathcal D_n)\). From the retained MCMC draws, this probability is estimated directly by
\[
\widehat{\Pr}(\theta_j>0\mid\mathcal D_n)
=
\frac{1}{M}\sum_{m=1}^M I\{\theta_j^{(m)}>0\}.
\]
Complete posterior summaries and convergence diagnostics for both the constant and varying
dispersion fits are collected in the application appendix. Detailed Markov
chain Monte Carlo (MCMC) diagnostics and model fit checks for all three
applications are collected in the diagnostic appendices. The number of
retained MCMC draws, effective sample sizes (ESS), and elapsed computation
times are also presented in the application appendix.

To summarize overall fitted behavior from the varying-dispersion model, for each retained draw \(m=1,\ldots,M\) we compute posterior samples
\[
\bar\mu^{(m)}=\frac{1}{n}\sum_{i=1}^n \mu_i\{\beta^{(m)}\},
\qquad
\bar\phi^{(m)}=\frac{1}{n}\sum_{i=1}^n \phi_i\{\gamma^{(m)}\}.
\]
The posterior density plots below are based directly on these retained MCMC draws. For a constant-dispersion fit, the analogous draw is \(\phi_0^{(m)}=\exp\{\gamma_0^{(m)}\}-1\). These summaries make it possible for us to report posterior probabilities of overdispersion or underdispersion directly on the dispersion scale instead of using delta-method based asymptotic inference.

\subsection{Student attendance}

The Attendance data contain \(n=314\) students. The response is the number of days absent. The mean model includes sex, program track, and math score. Program has General, Academic, and Vocational levels. The data are widely used as an illustration of overdispersed count regression \citep{UCLAAttendance}.

Table~\ref{tab:att} shows the fit for the Bayesian AETEL varying dispersion. The posterior sign probabilities give strong evidence that the male, Academic, Vocational, and math mean coefficients are negative: \(\Pr(\beta_{\rm male}>0\mid\mathcal D_n)=0.0219\), \(\Pr(\beta_{\rm Academic}>0\mid\mathcal D_n)=0.0012\), \(\Pr(\beta_{\rm Vocational}>0\mid\mathcal D_n)=0\), and \(\Pr(\beta_{\rm math}>0\mid\mathcal D_n)=0.0005\). A 10 point increase in math score corresponds to roughly 7 percent fewer expected absences using the posterior median coefficient from the retained draws. For dispersion, the Academic and Vocational effects are strongly positive, with posterior probabilities 0.9999 and 0.9997 of being above zero, whereas the sex and math dispersion effects have probabilities 0.3875 and 0.4841 of being positive and therefore show no clear posterior direction. Thus, program track is related to both the expected number of absences and the amount of extra variation around that mean.

\begin{table}[H]
\centering
\caption{Attendance data: Bayesian AETEL posterior summary for varying dispersion.}
\label{tab:att}
\scriptsize
\setlength{\tabcolsep}{4.5pt}
\resizebox{\textwidth}{!}{%
\begin{tabular}{lrrrr}
\toprule
Parameter & $\Pr(\theta_j>0\mid\mathcal D_n)$ & SD & Median & 95\% HPD \\
\midrule
$\beta_0$ & 1.0000 & 0.1462 & 2.7590 & $[2.4732, 3.0439]$ \\
$\beta_{\rm male}$ & 0.0219 & 0.1179 & -0.2390 & $[-0.4700, -0.0097]$ \\
$\beta_{\rm Academic}$ & 0.0012 & 0.1343 & -0.4107 & $[-0.6744, -0.1487]$ \\
$\beta_{\rm Vocational}$ & 0.0000 & 0.1742 & -1.2608 & $[-1.6044, -0.9187]$ \\
$\beta_{\rm math}$ & 0.0005 & 0.0023 & -0.0073 & $[-0.0120, -0.0029]$ \\
$\gamma_0$ & 0.9999 & 0.1024 & 0.3426 & $[0.1449, 0.5473]$ \\
$\gamma_{\rm male}$ & 0.3875 & 0.1017 & -0.0294 & $[-0.2256, 0.1682]$ \\
$\gamma_{\rm Academic}$ & 0.9999 & 0.0916 & 0.3249 & $[0.1475, 0.5090]$ \\
$\gamma_{\rm Vocational}$ & 0.9997 & 0.1299 & 0.5505 & $[0.2996, 0.8062]$ \\
$\gamma_{\rm math}$ & 0.4841 & 0.0020 & -0.0001 & $[-0.0040, 0.0037]$ \\
\bottomrule
\end{tabular}%
}
\end{table}

The constant dispersion fit gives a posterior median of
\(\gamma_0=0.5690\), corresponding to a posterior median of
\(\phi_0=\exp(0.5690)-1\approx0.7665\).
Moreover,
\(\Pr(\phi_0>0\mid\mathcal D_n)=1.000\), giving strong posterior
evidence of overdispersion. The varying fit shows that a single
overdispersion number hides meaningful program differences. The mean
coefficients, however, remain stable across the two specifications.

\begin{figure}[H]
\centering
\projectgraphic[
  width=0.82\linewidth,
  trim=0 3cm 0 0,
  clip
]{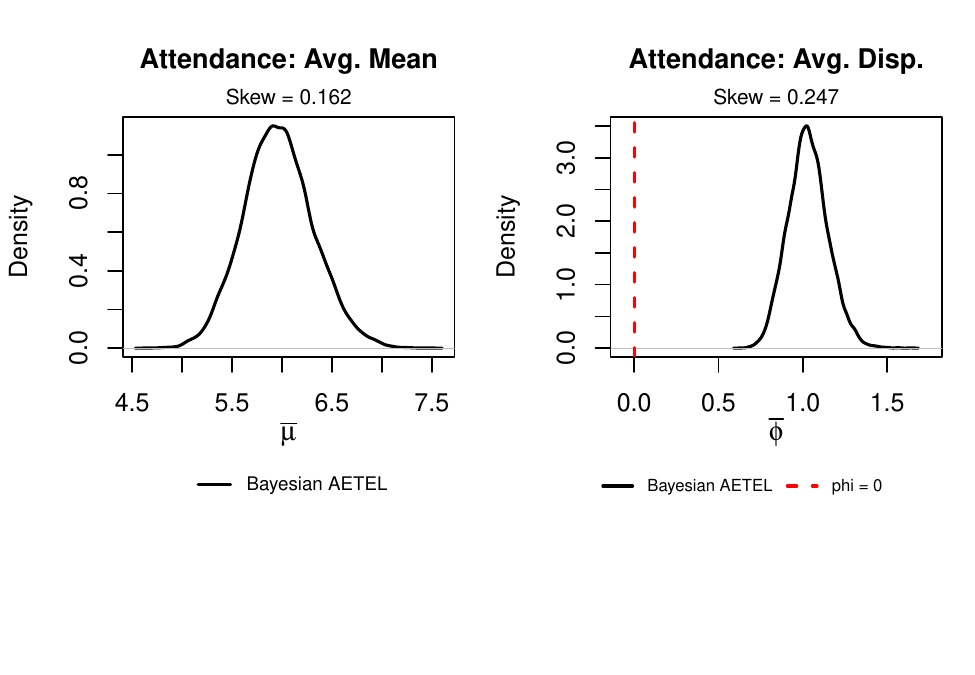}
\caption{Attendance data: posterior densities of the average mean
\(\bar\mu\) and average dispersion \(\bar\phi\) from the varying
dispersion fit. The dispersion panel summarizes the overall posterior
evidence for overdispersion relative to \(\bar\phi=0\).}
\label{fig:attendance_phi_posterior}
\end{figure}

Figure~\ref{fig:attendance_phi_posterior} gives a direct posterior summary of the
overall mean and dispersion under the varying dispersion model. The posterior
median of the average mean is about \(\bar\mu=5.96\) days absent. More
importantly for dispersion, the posterior median of \(\bar\phi\) is about
\(1.03\), and all retained draws lie above zero, giving
\(\Pr(\bar\phi>0\mid\mathcal D_n)=1.000\). Thus, the varying dispersion fit
provides strong posterior evidence of overall overdispersion in the Attendance
data. This agrees with the positive constant dispersion fit, and the
coefficient-specific results show that the amount of overdispersion varies
across program tracks.

\subsection{Takeover bids}

The Takeover Bids data record the number of bids received by United States firms facing tender offers between 1978 and 1985. Predictors include the opening bid premium, institutional holdings, legal and real asset restrictions, financial restrictions, regulatory action, white knight status, and firm size. Earlier analyses of these data have used parametric count models \citep{CameronJohansson1997}. We use a working sample of \(n=124\). The table labels BIDPREM, INSTHOLD, LEGLREST, REALREST, FINREST, REGULATN, WHTKNGHT, and logSIZE denote opening bid premium, institutional holdings, legal restrictions, real asset restrictions, financial restrictions, regulatory action, white knight status, and log firm size, respectively.

The varying dispersion posterior in Table~\ref{tab:takeover} shows strong posterior evidence that higher bid premium is associated with fewer expected bids, with \(\Pr(\beta_{\rm BIDPREM}>0\mid\mathcal D_n)=0.0181\), while white knight status and firm size are associated with more expected bids, with posterior probabilities 1.0000 and 0.9959 of positive effects. The dispersion results add information that is not visible in the mean model. Real asset restrictions and white knight status have strongly positive dispersion effects, with \(\Pr(\gamma_{\rm REALREST}>0\mid\mathcal D_n)=0.9951\) and \(\Pr(\gamma_{\rm WHTKNGHT}>0\mid\mathcal D_n)=0.9994\). Using the posterior median coefficients from the retained draws, these effects multiply \(1+\phi\) by \(\exp(0.6041)\approx1.83\) and \(\exp(0.5391)\approx1.71\), respectively, holding the other predictors fixed.

\begin{table}[H]
\centering
\caption{Takeover Bids data: Bayesian AETEL posterior summary for varying dispersion.}
\label{tab:takeover}
\scriptsize
\setlength{\tabcolsep}{4.5pt}
\resizebox{\textwidth}{!}{%
\begin{tabular}{lrrrr}
\toprule
Parameter & $\Pr(\theta_j>0\mid\mathcal D_n)$ & SD & Median & 95\% HPD \\
\midrule
$\beta_0$ & 0.9963 & 0.3861 & 1.0194 & $[0.2476, 1.7736]$ \\
$\beta_{\rm BIDPREM}$ & 0.0181 & 0.2758 & -0.5733 & $[-1.1114, -0.0281]$ \\
$\beta_{\rm INSTHOLD}$ & 0.0465 & 0.3211 & -0.5260 & $[-1.1659, 0.1072]$ \\
$\beta_{\rm LEGLREST}$ & 0.9595 & 0.1154 & 0.2086 & $[-0.0194, 0.4369]$ \\
$\beta_{\rm REALREST}$ & 0.0844 & 0.1743 & -0.2347 & $[-0.5771, 0.1065]$ \\
$\beta_{\rm FINREST}$ & 0.4014 & 0.2293 & -0.0577 & $[-0.5135, 0.3892]$ \\
$\beta_{\rm REGULATN}$ & 0.3622 & 0.1382 & -0.0482 & $[-0.3182, 0.2184]$ \\
$\beta_{\rm WHTKNGHT}$ & 1.0000 & 0.1146 & 0.5161 & $[0.2983, 0.7456]$ \\
$\beta_{\rm logSIZE}$ & 0.9959 & 0.0663 & 0.1612 & $[0.0321, 0.2960]$ \\
$\gamma_0$ & 0.4721 & 0.5595 & -0.0383 & $[-1.1232, 1.0480]$ \\
$\gamma_{\rm BIDPREM}$ & 0.0598 & 0.3828 & -0.5745 & $[-1.3380, 0.1620]$ \\
$\gamma_{\rm INSTHOLD}$ & 0.4814 & 0.4462 & -0.0209 & $[-0.8998, 0.8583]$ \\
$\gamma_{\rm LEGLREST}$ & 0.6033 & 0.1769 & 0.0454 & $[-0.2985, 0.3957]$ \\
$\gamma_{\rm REALREST}$ & 0.9951 & 0.2174 & 0.6041 & $[0.1798, 1.0314]$ \\
$\gamma_{\rm FINREST}$ & 0.8971 & 0.2059 & 0.2366 & $[-0.1603, 0.6538]$ \\
$\gamma_{\rm REGULATN}$ & 0.5189 & 0.1818 & 0.0066 & $[-0.3529, 0.3667]$ \\
$\gamma_{\rm WHTKNGHT}$ & 0.9994 & 0.1797 & 0.5391 & $[0.1889, 0.8925]$ \\
$\gamma_{\rm logSIZE}$ & 0.8365 & 0.0628 & 0.0545 & $[-0.0645, 0.1832]$ \\
\bottomrule
\end{tabular}%
}
\end{table}

The constant dispersion fit is close to equidispersion. On the dispersion scale, the posterior median is \(\phi_0=-0.0240\), with \(\Pr(\phi_0<0\mid\mathcal D_n)=0.6140\) and \(\Pr(\phi_0>0\mid\mathcal D_n)=0.3860\), so there is no strong posterior evidence for either direction under the constant model. The varying model shows why that single dispersion summary can be misleading because different firm characteristics push the conditional spread in different directions.

\begin{figure}[H]
\centering
\projectgraphic[width=0.82\linewidth]{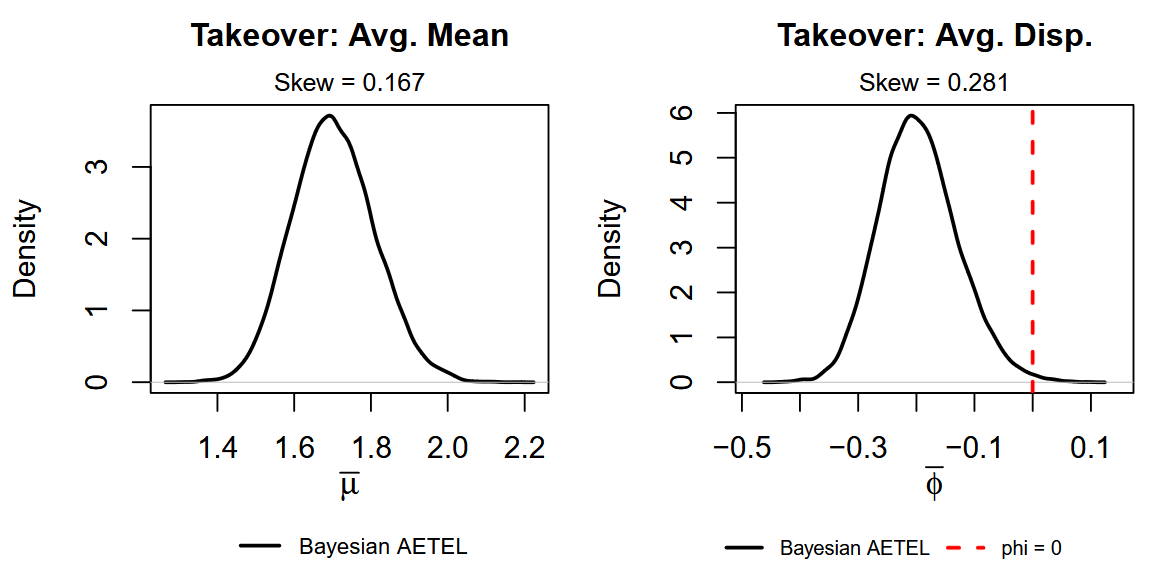}
\caption{Takeover Bids data: posterior densities of the average mean \(\bar\mu\) and average dispersion \(\bar\phi\) from the varying dispersion fit. The dispersion panel summarizes the overall posterior evidence relative to \(\bar\phi=0\).}
\label{fig:takeover_phi_posterior}
\end{figure}

Figure~\ref{fig:takeover_phi_posterior} shows that the posterior median of the
average mean is about \(\bar\mu=1.70\) bids. The posterior median of the average
dispersion is about \(\bar\phi=-0.199\), with
\(\Pr(\bar\phi<0\mid\mathcal D_n)=0.9946\) and
\(\Pr(\bar\phi>0\mid\mathcal D_n)=0.0054\). Hence, the varying dispersion fit
gives strong posterior evidence of overall underdispersion. This does not
conflict with the constant dispersion fit being close to equidispersion. The
constant model summarizes the spread with one number, whereas the varying model
allows firm characteristics to change dispersion across observations. The
positive dispersion effects for real asset restrictions and white knight
status therefore describe local changes around an overall average that is
below zero.

\subsection{Cotton bolls}

The Cotton Bolls experiment counts bolls on plants observed at five stages: vegetative, bud, blossom, fig, and boll. The design contains \(n=125\) observations, with \(25\) at each stage, together with a defoliation measure. The mean model includes stage, centered defoliation, and their interactions. Similar count data have motivated flexible dispersion models in agricultural applications \citep{ZevianiEtAl2014}.

The mean structure in Table~\ref{tab:cotton} is stable relative to the constant dispersion fit. Defoliation has a negative coefficient in the vegetative reference stage, with \(\Pr(\beta_{\rm defc}>0\mid\mathcal D_n)=0\), and several stage interactions change the slope. The dispersion block is more complex. The blossom by defoliation interaction is strongly negative, the fig by defoliation interaction is strongly positive, and the boll by defoliation interaction is strongly negative, with posterior probabilities of being positive equal to 0.0000, 0.9958, and 0.0001, respectively. The main fig dispersion coefficient has \(\Pr(\gamma_{\rm fig}>0\mid\mathcal D_n)=0.9771\), although its 95 percent HPD interval still contains zero, so the evidence for this main effect is less decisive than for the three interactions.

\begin{table}[H]
\centering
\caption{Cotton Bolls data: Bayesian AETEL posterior summary for varying dispersion.}
\label{tab:cotton}
\scriptsize
\setlength{\tabcolsep}{4.5pt}
\resizebox{\textwidth}{!}{%
\begin{tabular}{lrrrr}
\toprule
Parameter & $\Pr(\theta_j>0\mid\mathcal D_n)$ & SD & Median & 95\% HPD \\
\midrule
$\beta_0$ & 1.0000 & 0.0254 & 2.1166 & $[2.0668, 2.1657]$ \\
$\beta_{\rm bud}$ & 0.8387 & 0.0379 & 0.0367 & $[-0.0361, 0.1113]$ \\
$\beta_{\rm blossom}$ & 0.0000 & 0.0496 & -0.2960 & $[-0.3928, -0.1991]$ \\
$\beta_{\rm fig}$ & 0.0000 & 0.0598 & -0.2391 & $[-0.3564, -0.1215]$ \\
$\beta_{\rm boll}$ & 0.9971 & 0.0380 & 0.1003 & $[0.0262, 0.1746]$ \\
$\beta_{\rm defc}$ & 0.0000 & 0.0778 & -0.3786 & $[-0.5301, -0.2267]$ \\
$\beta_{\rm bud:defc}$ & 0.9878 & 0.1104 & 0.2460 & $[0.0312, 0.4638]$ \\
$\beta_{\rm blossom:defc}$ & 0.0715 & 0.1428 & -0.2096 & $[-0.4894, 0.0698]$ \\
$\beta_{\rm fig:defc}$ & 0.0096 & 0.1552 & -0.3661 & $[-0.6685, -0.0617]$ \\
$\beta_{\rm boll:defc}$ & 0.9788 & 0.1135 & 0.2284 & $[0.0073, 0.4514]$ \\
$\gamma_0$ & 0.0000 & 0.0062 & -0.1048 & $[-0.1164, -0.0923]$ \\
$\gamma_{\rm bud}$ & 0.8307 & 0.0097 & 0.0094 & $[-0.0098, 0.0285]$ \\
$\gamma_{\rm blossom}$ & 0.0811 & 0.0215 & -0.0271 & $[-0.0698, 0.0137]$ \\
$\gamma_{\rm fig}$ & 0.9771 & 0.0275 & 0.0516 & $[-0.0018, 0.1058]$ \\
$\gamma_{\rm boll}$ & 0.9634 & 0.0093 & 0.0165 & $[-0.0014, 0.0346]$ \\
$\gamma_{\rm defc}$ & 0.8712 & 0.0167 & 0.0182 & $[-0.0138, 0.0510]$ \\
$\gamma_{\rm bud:defc}$ & 0.1183 & 0.0240 & -0.0276 & $[-0.0744, 0.0187]$ \\
$\gamma_{\rm blossom:defc}$ & 0.0000 & 0.0827 & -0.2268 & $[-0.3972, -0.0756]$ \\
$\gamma_{\rm fig:defc}$ & 0.9958 & 0.0835 & 0.2058 & $[0.0395, 0.3741]$ \\
$\gamma_{\rm boll:defc}$ & 0.0001 & 0.0229 & -0.0748 & $[-0.1206, -0.0313]$ \\
\bottomrule
\end{tabular}%
}
\end{table}

For the constant dispersion model, the posterior median on the dispersion scale is \(\phi_0=-0.0909\), with \(\Pr(\phi_0<0\mid\mathcal D_n)=1.000\), so the counts are mildly underdispersed on average with strong posterior evidence for the direction. The varying fit shows that the amount of underdispersion depends on stage and defoliation. Because each stage contains only 25 observations and the model has many interactions, uncertainty is larger for several of these local effects than in the other two examples.

\begin{figure}[H]
\centering
\projectgraphic[width=0.82\linewidth]{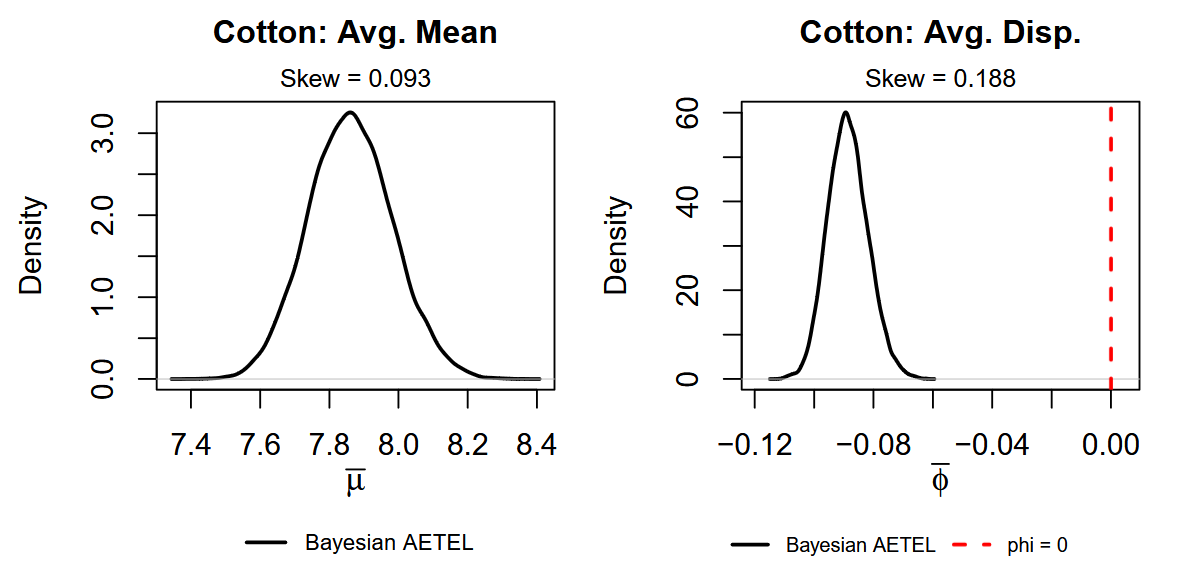}
\caption{Cotton Bolls data: posterior densities of the average mean \(\bar\mu\) and average dispersion \(\bar\phi\) from the varying dispersion fit. The dispersion panel summarizes the overall posterior evidence for underdispersion relative to \(\bar\phi=0\).}
\label{fig:cotton_phi_posterior}
\end{figure}

Figure~\ref{fig:cotton_phi_posterior} gives a posterior median of the average
mean of about \(\bar\mu=7.86\) bolls. The posterior for average dispersion is
concentrated below zero, with posterior median about \(\bar\phi=-0.089\) and
\(\Pr(\bar\phi<0\mid\mathcal D_n)=1.000\). This gives strong posterior evidence
of overall underdispersion and agrees closely with the constant dispersion
summary \(\phi_0\approx-0.091\). The varying dispersion coefficients then show
how this overall underdispersion changes with stage and defoliation.

\subsection{Relation to frequentist GMM}

The corresponding frequentist zero altered regression model was studied using
GMM by \citet{GhoshBoakyeSellers2026}. The Bayesian AETEL and frequentist GMM
analyses use the same first two factorial moment restrictions, but differ in how uncertainty and interval
estimates are obtained. Frequentist GMM gives a minimum distance estimator
together with sandwich-based standard errors and confidence intervals obtained
from the asymptotic sampling distribution of the estimator. Bayesian AETEL uses
the moment restrictions to form an empirical likelihood criterion and samples
the resulting AETEL pseudo-posterior distribution. Posterior medians, posterior
standard deviations, credible intervals, including the HPD intervals reported
in this paper, and direct posterior summaries for nonlinear functions such as
\(\phi_i=\exp(x_i\trans\gamma)-1\) are obtained from the retained MCMC draws.

The empirical conclusions are also closely aligned with the frequentist GMM
results obtained by \citet{GhoshBoakyeSellers2026}. Although the two approaches
construct uncertainty differently, the estimated mean and dispersion effects
lead to similar substantive conclusions. The present analysis focuses on the
Bayesian AETEL posterior summaries and their corresponding uncertainty
quantification.

The application results also show what the Bayesian analysis adds when the
frequentist GMM estimates are already available. The posterior draws give
probability statements directly for quantities that are scientifically easy to
interpret. For each regression coefficient, Tables~\ref{tab:att},
\ref{tab:takeover}, and~\ref{tab:cotton} report
\(\Pr(\theta_j>0\mid\mathcal D_n)\), so the posterior evidence for the sign of
an effect is read directly rather than inferred only from a local normal
approximation. The same idea is especially useful for the nonlinear dispersion
functional. The Attendance analysis gives
\(\Pr(\bar\phi>0\mid\mathcal D_n)=1.000\), and the Takeover Bids and Cotton
Bolls analyses give \(\Pr(\bar\phi<0\mid\mathcal D_n)=0.9946\) and \(1.000\),
respectively. These posterior probabilities and the corresponding density plots
summarize uncertainty in overall dispersion on the scale of direct scientific
interest. They complement the frequentist GMM analysis.
Both procedures use the same moment restrictions, close point estimates
are expected, while Bayesian AETEL additionally provides the full AETEL
pseudo-posterior distribution and direct probability statements for coefficient
signs and nonlinear functions of the parameters. In particular, uncertainty for
\(\bar\phi\) is obtained by transforming each retained posterior draw without using a separate delta-method approximation. This is useful when the nonlinear transformation is
not well summarized by a local Gaussian approximation.
Under the regularity conditions in Section~\ref{sec:theory}, the two procedures use the same first-order information, so close results are expected when the sample information is strong and prior regularization is weak. The asymptotic distribution in Section~\ref{sec:theory} describes the large-sample behavior of the Bayesian AETEL posterior but is not used to construct the Bayesian intervals. The simulation results in Tables~\ref{tab:sim_zip} and~\ref{tab:sim_cmp} illustrate this first-order agreement.

The Bayesian approach is not claimed to be uniformly preferable. Its advantage is that inference is based on sampling the full AETEL pseudo-posterior but not summarizing uncertainty only through the asymptotic sampling distribution of an optimized criterion. The cost is substantially greater computation time because the tilting equation must be solved repeatedly. The delayed acceptance construction in Section~\ref{sec:aetel} is used to reduce that computational burden by screening proposals before an exact AETEL evaluation is required.

\section{Discussion and Conclusion}
\label{sec:conclusion}

We developed a Bayesian semiparametric regression framework for dispersed count data using the first two factorial moments of a zero altered model \citep{GhoshKim2007}. The moment based formulation does not require a complete parametric probability model for the counts. The Bayesian approach also provides posterior uncertainty for both the mean and dispersion effects.

The adjusted exponentially tilted empirical likelihood is useful for two reasons. The large-sample argument uses existing Bayesian ETEL theory, and the model specific calculations verify the moment envelopes and Jacobian structure for the zero altered factorial moments. Statistically, it converts the estimating equations into a likelihood-type criterion that can be combined with a prior. Numerically, the extra artificial moment reduces failures that occur when the observed moment vectors cannot support a weighted average equal to zero. The two block delayed acceptance sampler addresses the remaining computational cost by screening proposals with a frozen surrogate and evaluating the exact AETEL criterion only after a proposal passes the first stage. The second-stage correction retains the exact adjusted AETEL pseudo-posterior as the invariant target.

The simulations show that the procedure can recover mean and dispersion coefficients under both overdispersion and underdispersion. Empirical variability and reported posterior uncertainty were close, and interval coverage was near 95 percent. The three applications show why modeling dispersion as a regression component can be very important. Program track changed dispersion in student absences, firm actions changed dispersion in takeover bids, and stage by defoliation interactions changed dispersion in cotton boll counts. A single dispersion parameter would average over these differences. The posterior dispersion densities make these conclusions especially transparent: the posterior mass for average dispersion is above zero for Attendance and below zero for Takeover Bids and Cotton Bolls. Together with the coefficient sign probabilities, these summaries illustrate the additional probability statements available from the Bayesian formulation and remains consistent with the corresponding frequentist GMM conclusions.

Several directions for future work are possible. One important direction is the high dimensional case, where the numbers of mean and dispersion covariates may increase with the sample size. The mean and dispersion blocks could also use different covariate sets and different dimensions rather than the common design used here. Variable selection or shrinkage methods could then be used to identify the important mean and dispersion effects. Sparse Gaussian priors, continuous shrinkage priors, or spike and slab constructions could also prevent a complex dispersion model from using too much local information \citep{GeorgeMcCulloch1993,ParkCasella2008,IshwaranRao2005}. A second direction is correlated count data, including clustered or repeated measurements, where observations within the same cluster may not be independent. Extending the moment conditions and their covariance structure to account for within cluster dependence would make the method useful for longitudinal and clustered studies \citep{LiangZeger1986}. Another direction is flexible instrument choice and overidentified moment systems, where Bayesian model comparison could help assess competing moment restrictions \citep{Hansen1982,NeweySmith2004,ChibShinSimoni2018}. Finally, the computational framework can be improved further by studying alternative surrogate constructions, MCMC proposal schemes, and block proposal scaling as the dimension grows.

A full count likelihood is not necessary when the scientific structure can be expressed through reliable low order moments. When we combine zero altered moments with adjusted exponential tilting we obtain a Bayesian empirical-likelihood procedure that remains flexible across both directions of dispersion and can reveal how uncertainty changes with observed covariates.

\appendix

\section{Proof of the Moment Identities}
\label{app:moments}

This appendix gives the algebra behind the moment equations used throughout the paper.

\begin{proof}[Proof of the moment identities]
From \eqref{eq:za_pmf}, the zero altered mean is \(\mu=\omega\mu_0\). The variance identity for the zero altered construction is \(\Var(Y)=\omega\Var_{f_0}(U)+(1-\omega)\mu^2/\omega\). Because the baseline distribution is equidispersed, \(\Var_{f_0}(U)=\mu_0\), so \(\Var(Y)=\omega\mu_0+(1-\omega)\mu^2/\omega=\mu+(1-\omega)\mu^2/\omega\). This proves \eqref{eq:qvf} with \(\phi=(1-\omega)/\omega\). Next, use \(\E\{Y(Y-1)\}=\Var(Y)+\{\E(Y)\}^2-\E(Y)\). Substituting \(\E(Y)=\mu\) and \eqref{eq:qvf} which gives \(\E\{Y(Y-1)\}=(1+\phi)\mu^2\).
Conditioning on the regressors and substituting the regression links in \eqref{eq:links} yields the two conditional moment restrictions collected in \eqref{eq:moment}. \qedhere
\end{proof}

\section{Verification of Regularity Conditions for the Zero Altered Regression Moments}
\label{app:theory}

This appendix gives the detailed model-specific verification used in Lemma~\ref{prop:zamregularity} and Theorem~\ref{thm:bvm}. The general Bayesian ETEL posterior concentration and local expansion conditions are taken from \citet{ChibShinSimoni2018}. Assumptions~\ref{ass:zamregular}--\ref{ass:identinfo} provide the independent and identically distributed structure, compactness, identification, positive definiteness of \(S_0\), and the prior condition. We verify below the model-specific conditions: a square-integrable envelope for the moment vector, an integrable envelope for its first derivative, and full column rank of the population Jacobian \(G_0\). The artificial AETEL moment is first-order negligible when \(a_n=o(\sqrt n)\), as used in the main text.

Choose \(K_x<\infty\) such that \(\|x_i\|\le K_x\) almost surely. Compactness of the parameter set implies that the linear predictors range over compact sets. Under Assumption~\ref{ass:zamregular}, finite constants \(M_g,M_h,M_{g'},M_{h'}\) therefore exist such that \(|g|\le M_g\), \(|h|\le M_h\), \(|g'|\le M_{g'}\), and \(|h'|\le M_{h'}\) uniformly over the relevant parameter set.

Denote the two blocks of \eqref{eq:moment} by
\(m_{1i}(\theta)=x_i\{Y_i-g(x_i\trans\beta)\}\) and
\(m_{2i}(\theta)=x_i[Y_i(Y_i-1)-g(x_i\trans\beta)^2h(x_i\trans\gamma)]\). Then
\[
\|m_{1i}(\theta)\|\le K_x(|Y_i|+M_g),
\qquad
\|m_{2i}(\theta)\|\le K_x\{Y_i^2+|Y_i|+M_g^2M_h\}.
\]
Hence the complete moment vector is bounded by a constant times \(1+|Y_i|+Y_i^2\). Since \(\E(Y_i^4)<\infty\), its squared envelope is integrable and \(m_i(\theta)\) has a square integrable envelope uniformly over the compact parameter set.

The derivative blocks are
\[
\frac{\partial m_{1i}}{\partial\beta\trans}
=-x_ix_i\trans g'(x_i\trans\beta),
\qquad
\frac{\partial m_{1i}}{\partial\gamma\trans}=0,
\]
and
\[
\frac{\partial m_{2i}}{\partial\beta\trans}
=-2x_ix_i\trans g(x_i\trans\beta)g'(x_i\trans\beta)h(x_i\trans\gamma),
\qquad
\frac{\partial m_{2i}}{\partial\gamma\trans}
=-x_ix_i\trans g(x_i\trans\beta)^2h'(x_i\trans\gamma).
\]
These blocks are uniformly bounded because \(x_i\), the links, and their first
derivatives are bounded on the relevant compact sets. Thus the derivative
has an integrable envelope. Under these envelope and continuity conditions,
the uniform law of large numbers applies to the moment vector and its
derivative, and the square integrability gives the multivariate central limit
theorem for \(n^{-1/2}\sum_{i=1}^n m_i(\theta_0)\); see, for example,
\citet{NeweyMcFadden1994}.

For the log links used in the simulations and applications, \(g'=g=\mu_i\) and \(h'=h=1+\phi_i\). Taking expectations of the derivative blocks at \(\theta_0\) gives
\[
G_0
=
-\begin{pmatrix}
\E(\mu_i x_ix_i\trans) & 0\\
2\E\{\mu_i^2(1+\phi_i)x_ix_i\trans\} &
\E\{\mu_i^2(1+\phi_i)x_ix_i\trans\}
\end{pmatrix}
=
-\begin{pmatrix}
A_\beta&0\\
2A_\gamma&A_\gamma
\end{pmatrix}.
\]
This matrix is block lower triangular. Under Assumption~\ref{ass:identinfo}, both diagonal blocks are positive definite, so \(G_0\) has full column rank.

\section{Numerical Evaluation and Posterior Simulation Details}
\label{app:computation}

For fixed \(\theta\), the adjusted moment dimension is \(d=2p\) for the full varying dispersion model and \(d=p+1\) for the reduced constant dispersion fits. The tilting equation is \(\bar m_w(\lambda;\theta)=\sum_{i=1}^{N}p_i(\theta,\lambda)m_i^a(\theta)=0\), and its Jacobian is the weighted covariance matrix \(H(\lambda;\theta)=\sum_{i=1}^{N}p_i(\theta,\lambda)C_i(\lambda;\theta)C_i(\lambda;\theta)\trans\). Starting from a warm value of \(\lambda\), the implementation updates \(\lambda_{\rm new}=\lambda-tH(\lambda;\theta)^{-1}\bar m_w(\lambda;\theta)\), where \(t\in(0,1]\) is reduced by backtracking until the norm of the tilting equation decreases. Eigenvalues of \(H(\lambda;\theta)\) smaller than a numerical floor are replaced by that floor before inversion \citep{NocedalWright2006}.

The delayed acceptance sampler begins with an exact pilot phase \citep{ChristenFox2005}. During this phase, the mean and dispersion blocks are updated with exact Metropolis steps and each exact log AETEL value is added to a cache. After the pilot, the cache is frozen. For a new parameter value, the surrogate log criterion is computed by inverse distance weighting over nearby cached states. The same prior contribution is then added to obtain the surrogate log posterior used at the first stage.

For the simulation study, each chain used 12,000 iterations, a burn in of 4,000 iterations, and thinning by 8. The first 3,000 iterations formed the exact pilot. The frozen surrogate used the 40 nearest cached states, so \(k=40\), with inverse distance power \(\kappa=1.5\) and a maximum cache size of \(N_c=5{,}000\). The Gaussian proposal scales were 1.0 for the mean block and 0.75 for the dispersion block. Four chains were run for the diagnostic fits. These values describe the simulation implementation.

For the real data analyses, the same two block delayed acceptance construction was used with data specific run lengths and tuning. The design matrices were standardized when needed for numerical stability and posterior draws were transformed back to the original covariate scale before reporting.

To prevent overflow in numerical exponentiation, we clip only extreme working linear predictors before applying the exponential. 

The Gaussian regularization used in the application can be written in precision form as \(P_\beta=\operatorname{diag}(s_{\beta,1}^{-2},\ldots,s_{\beta,p}^{-2})+r_\beta I\) and
\[
P_\gamma
=
\frac{X_\phi\trans X_\phi+r_\gamma I}{\tau_\gamma^2}
+
\operatorname{diag}(s_{\gamma,1}^{-2},\ldots,s_{\gamma,q}^{-2}),
\]
where \(X_\phi\) is the standardized dispersion working design, with \(q=p\) and \(X_\phi=X\) in the varying dispersion fits and \(q=1\) in the constant dispersion fits. On the standardized working scale, \(\beta\sim N(0,P_\beta^{-1})\) and \(\gamma\sim N(0,P_\gamma^{-1})\). Thus, the dispersion block includes a design scaled term whereas the mean block uses diagonal Gaussian precision. This difference was used in the computations and is kept here explicitly as finite sample regularization.

\section{Simulation Data Generation}
\label{app:simulation}

\subsection{Overdispersed zero inflated Poisson setting}

For each simulated covariate \(x\), define \(\mu=\exp(\beta_0+\beta_1x)\) and \(\phi=\exp(\gamma_0+\gamma_1x)-1\). In the overdispersed setting \(\phi>0\). Set \(\rho=\phi/(1+\phi)\) and \(\nu=\mu/(1-\rho)\).
Then generate \(Y=0\) with probability \(\rho\), and otherwise generate \(Y\sim\mathrm{Poisson}(\nu)\) \citep{Lambert1992}. This gives \(\E(Y\mid x)=\mu\) and \(\E\{Y(Y-1)\mid x\}=\mu^2(1+\phi)\).

\subsection{Underdispersed Conway Maxwell Poisson setting}

For the underdispersed setting, a Conway Maxwell Poisson distribution is calibrated to the same target moments \citep{ShmueliEtAl2005,SellersShmueli2010}. At each \(x\), numerical root finding chooses its two distribution parameters so that \(\E(Y\mid x)=\mu(x)\) and \(\E\{Y(Y-1)\mid x\}=\mu(x)^2\{1+\phi(x)\}\).
To reduce Monte Carlo cost, the distribution parameters are solved on a grid of 401 equally spaced \(x\) values in \([-1,1]\) and interpolated during data generation.

\setcounter{figure}{0}
\section{Additional Simulation Figures and Bayesian Diagnostics}
\label{app:simfigs}

\subsection{Overdispersed zero inflated Poisson setting}

In the simulation figures, FGMM denotes frequentist generalized method of moments and BAETEL denotes Bayesian adjusted exponentially tilted empirical likelihood.

\begin{figure}[H]
    \centering
    \projectgraphic[width=1.0\linewidth]{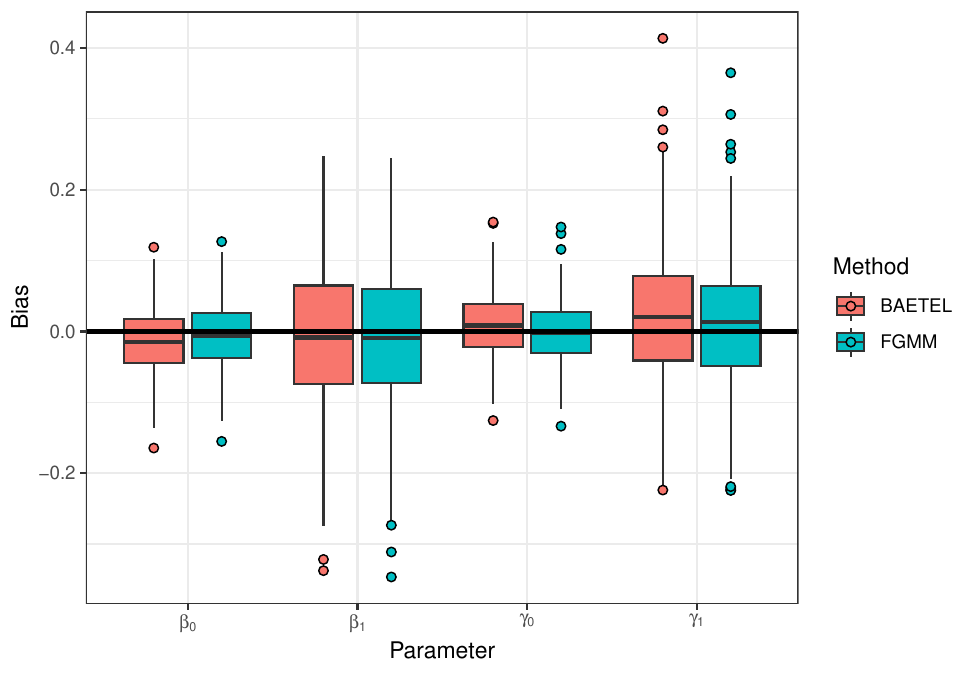}
    \caption{ZIP simulation: boxplots of estimation bias for the four regression coefficients.}
    \label{fig:zip_bias_box}
\end{figure}

\begin{figure}[H]
    \centering
    \begin{minipage}{0.48\linewidth}
        \centering
        \projectgraphic[width=\linewidth]{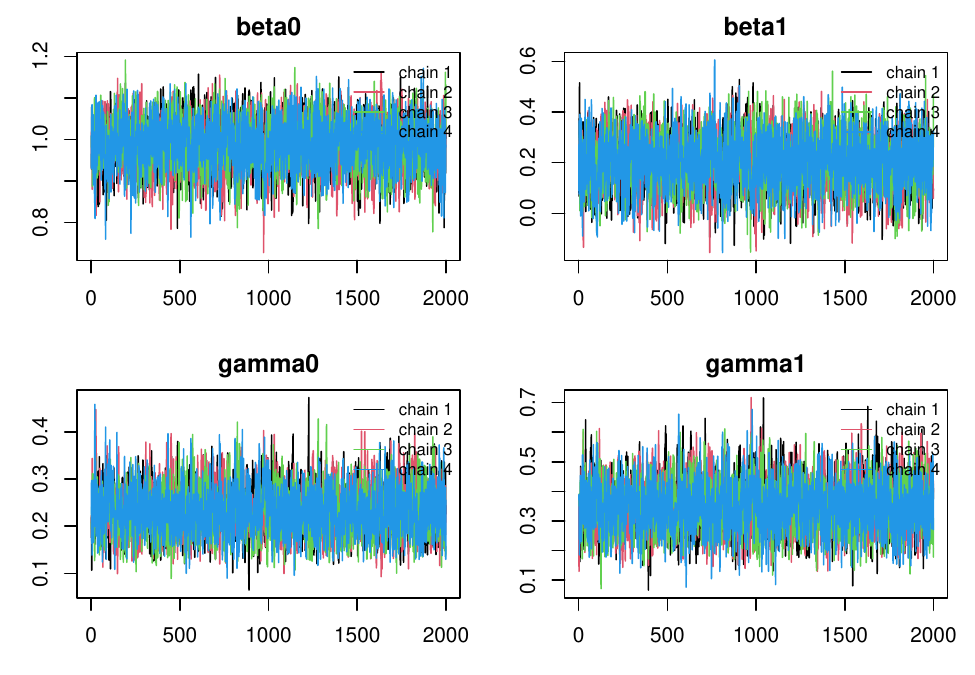}
    \end{minipage}
    \hfill
    \begin{minipage}{0.48\linewidth}
        \centering
        \projectgraphic[width=\linewidth]{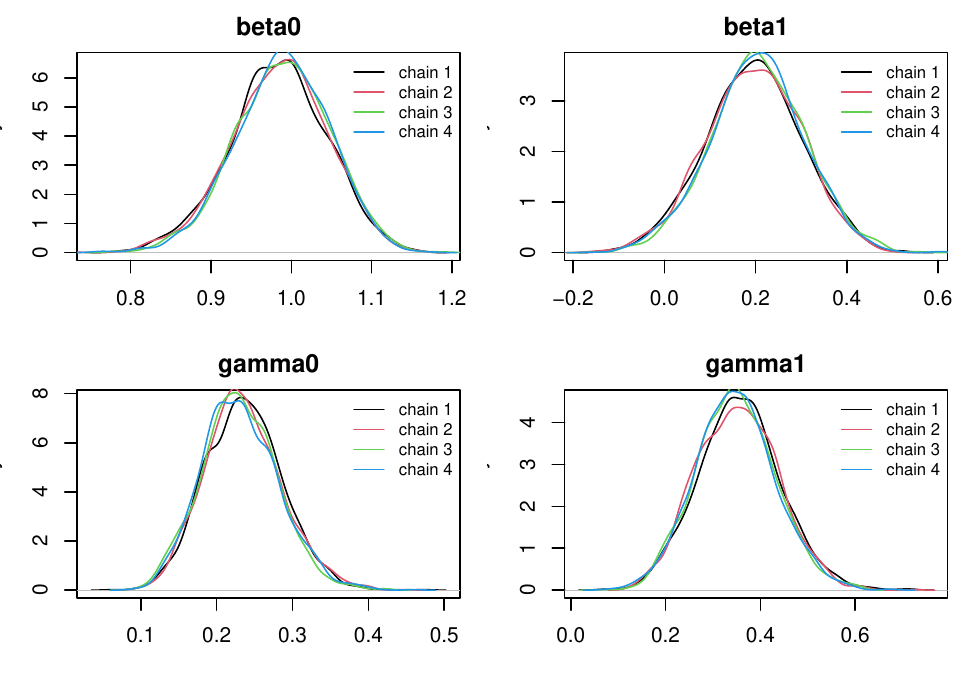}
    \end{minipage}
    \hfill
    \begin{minipage}{0.48\linewidth}
        \centering
        \projectgraphic[width=\linewidth]{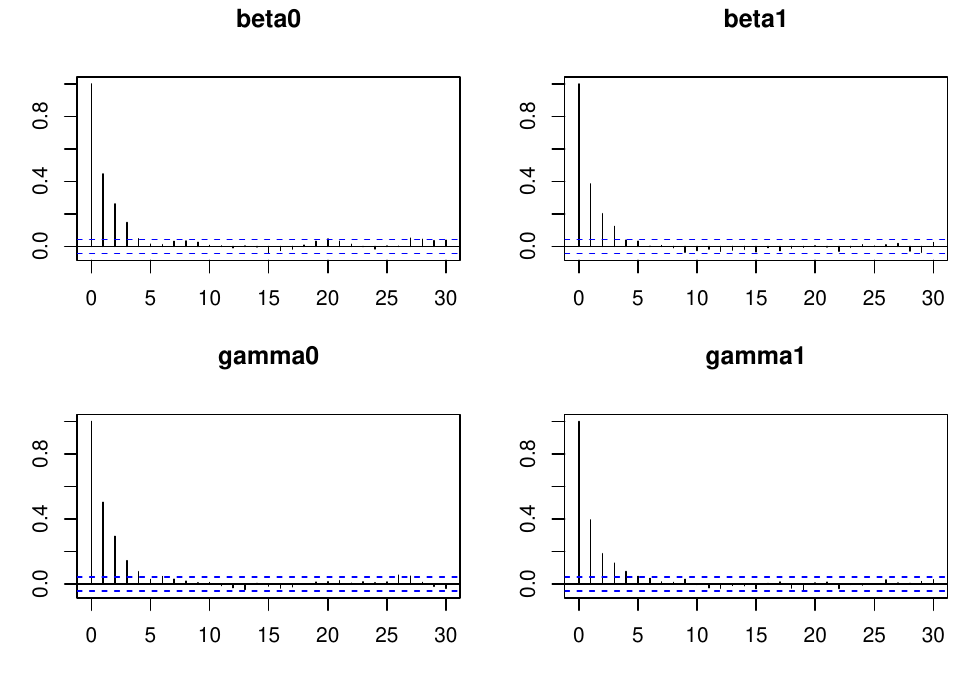}
    \end{minipage}
    \caption{Bayesian AETEL diagnostics for the ZIP simulation setting.}
    \label{fig:zip_bayesian_diagnostics}
\end{figure}

Figure~\ref{fig:zip_bias_box} shows that the sampling distributions of the four coefficient estimates are centered close to zero, with no evidence of systematic distortion in either the mean or dispersion block. Figure~\ref{fig:zip_bayesian_diagnostics} shows stable posterior behavior for the simulation sampler. The chains mix well, the posterior densities are smooth and unimodal, and the autocorrelation decays quickly enough to support the effective sample sizes.

\subsection{Underdispersed Conway Maxwell Poisson setting}

\begin{figure}[H]
    \centering
    \projectgraphic[width=0.9\linewidth]{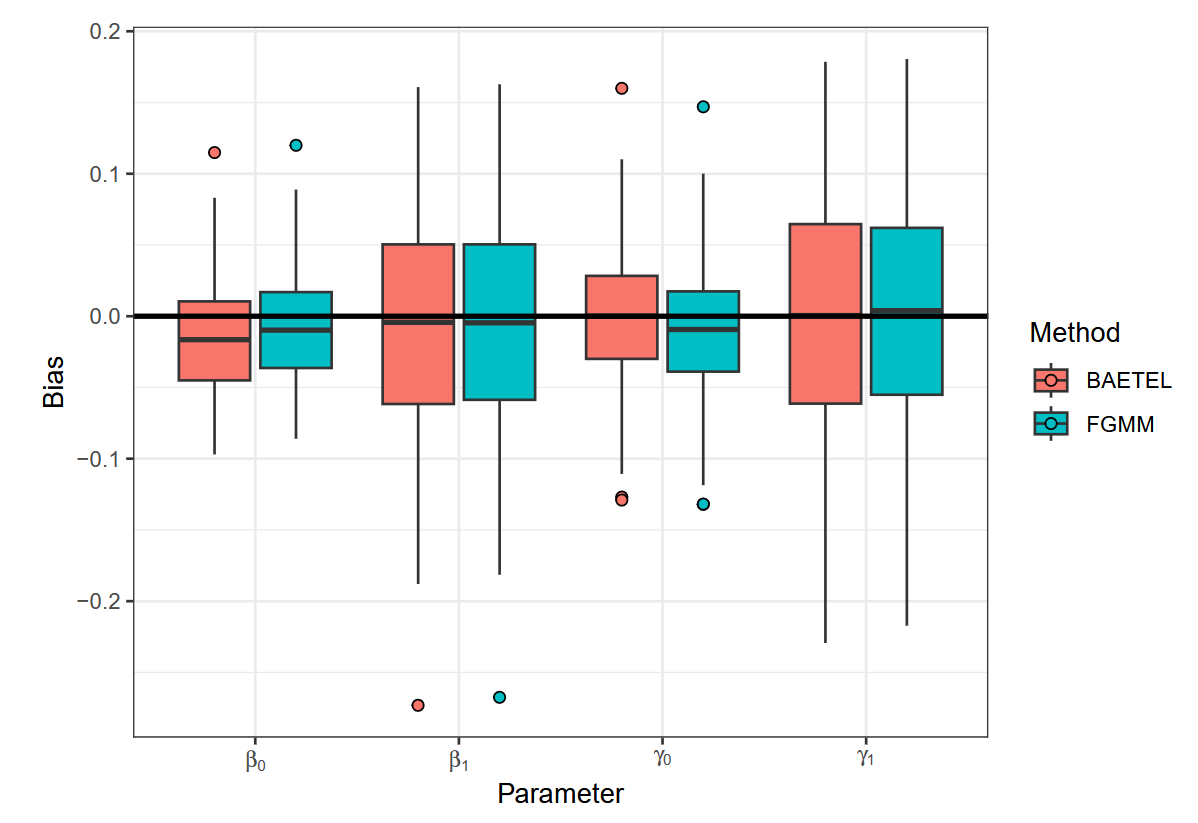}
    \caption{CMP simulation: boxplots of estimation bias for the four regression coefficients.}
    \label{fig:cmp_bias_box}
\end{figure}

\begin{figure}[H]
    \centering
    \begin{minipage}{0.72\linewidth}
        \centering
        \projectgraphic[width=\linewidth]{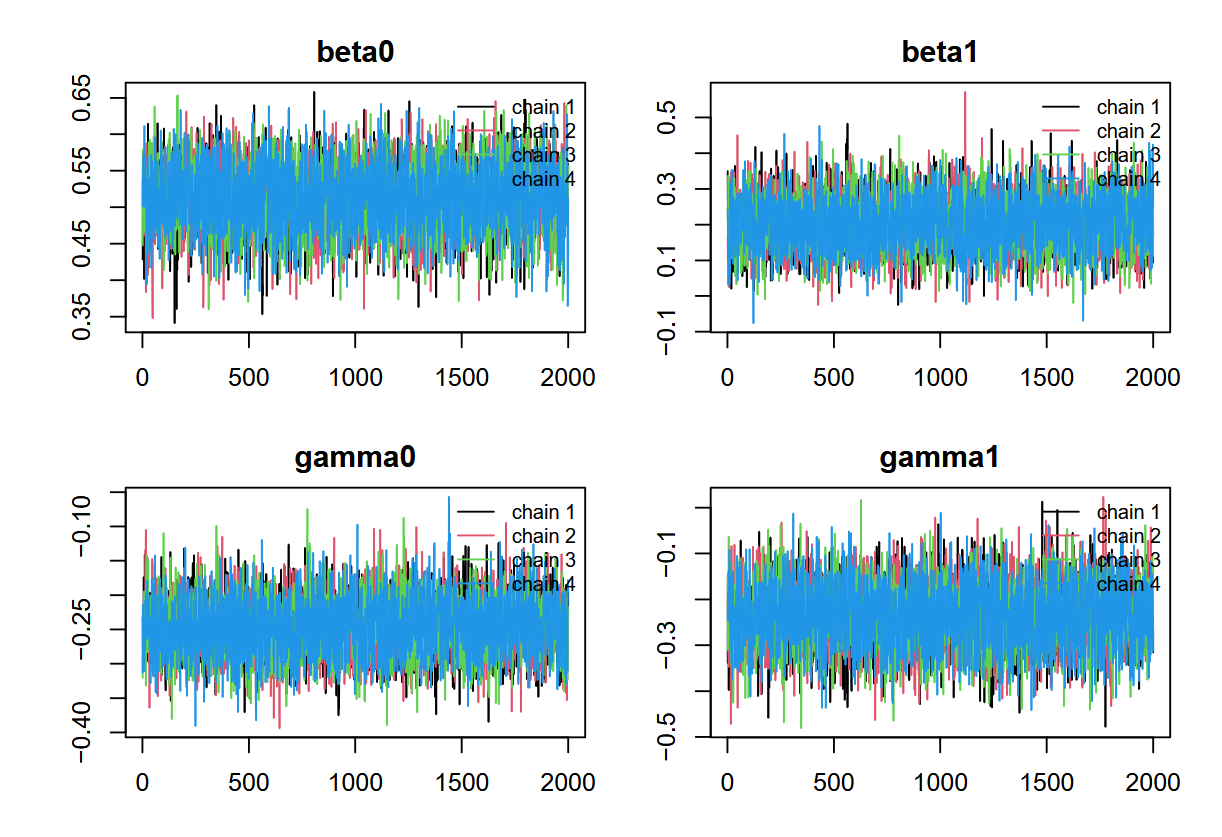}
    \end{minipage}
    \hfill
    \begin{minipage}{0.48\linewidth}
        \centering
        \projectgraphic[width=\linewidth]{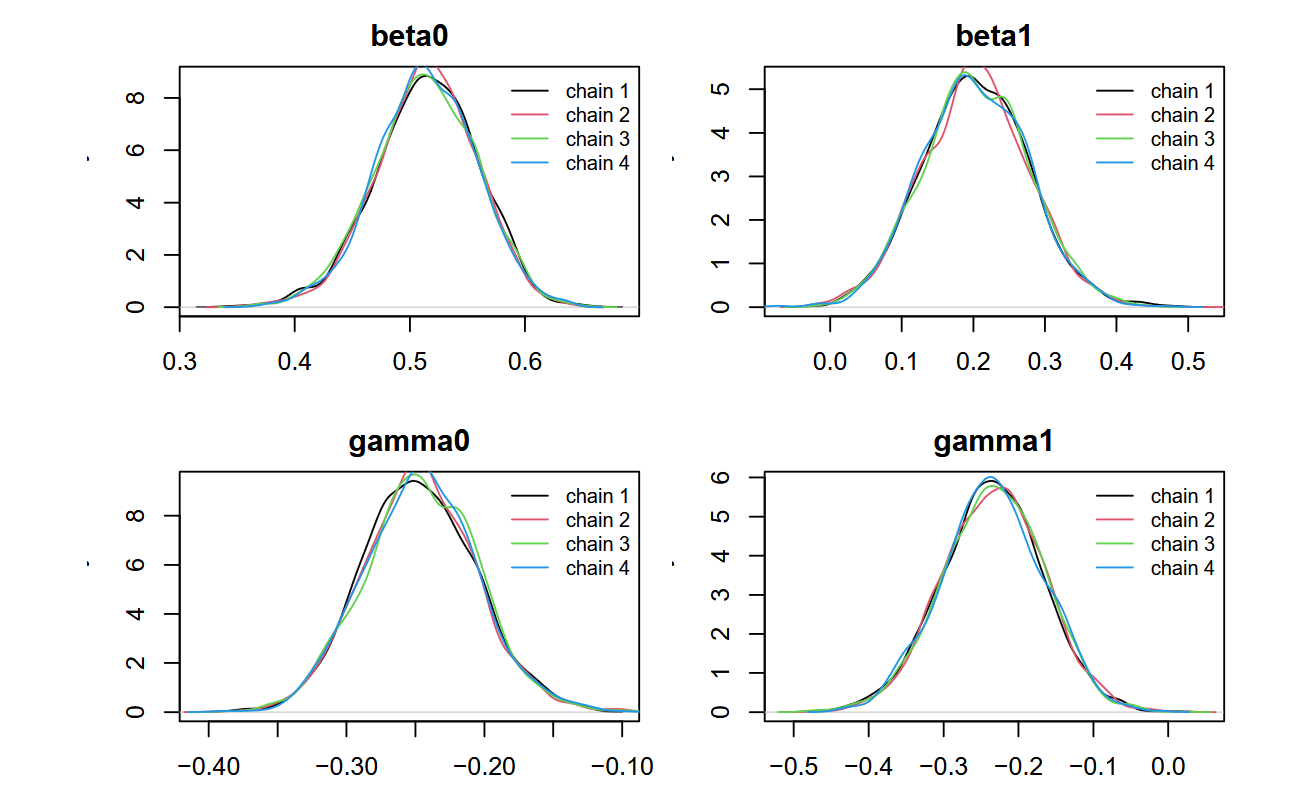}
    \end{minipage}
    \hfill
    \begin{minipage}{0.48\linewidth}
        \centering
        \projectgraphic[width=\linewidth]{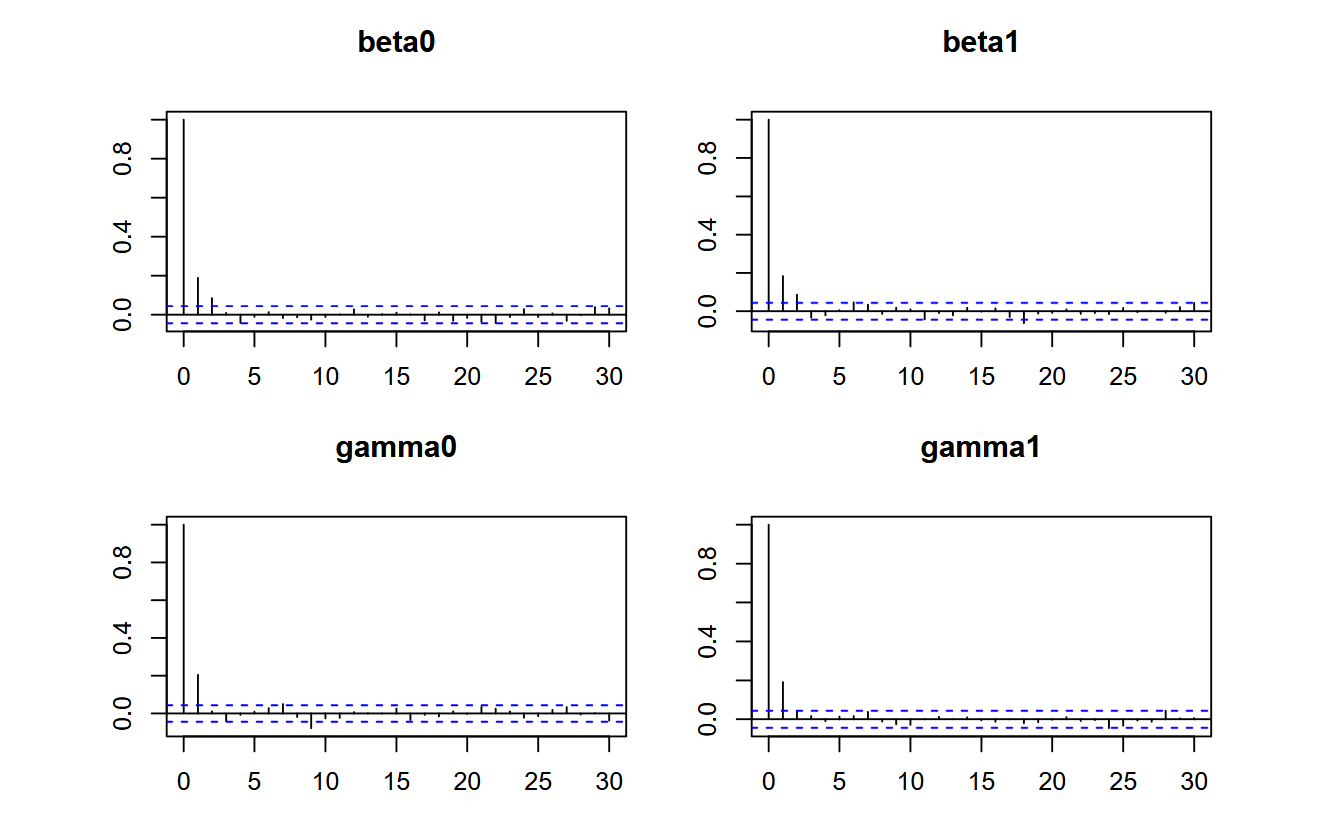}
    \end{minipage}
    \caption{Bayesian AETEL diagnostics for the CMP simulation setting.}
    \label{fig:cmp_bayesian_diagnostics}
\end{figure}

The CMP bias boxplots in Figure~\ref{fig:cmp_bias_box} tell the same information as the numerical summaries in Table~\ref{tab:sim_cmp}. The empirical bias remains small for all four coefficients even though the fitting method does not use the true parametric Conway Maxwell Poisson likelihood. The MCMC diagnostics in Figure~\ref{fig:cmp_bayesian_diagnostics} again indicate satisfactory mixing and stable posterior exploration.

\setcounter{figure}{0}
\section{Attendance MCMC Diagnostics}
\label{app:attendancediag}

In the diagnostic figures, ACF denotes autocorrelation function. The tables
have effective sample size (ESS), the potential scale reduction factor
\(\widehat R\), and its upper confidence limit (Upper CL). 

\begin{figure}[H]
\centering

\begin{minipage}[t]{0.62\linewidth}
\centering
\includegraphics[width=\linewidth]{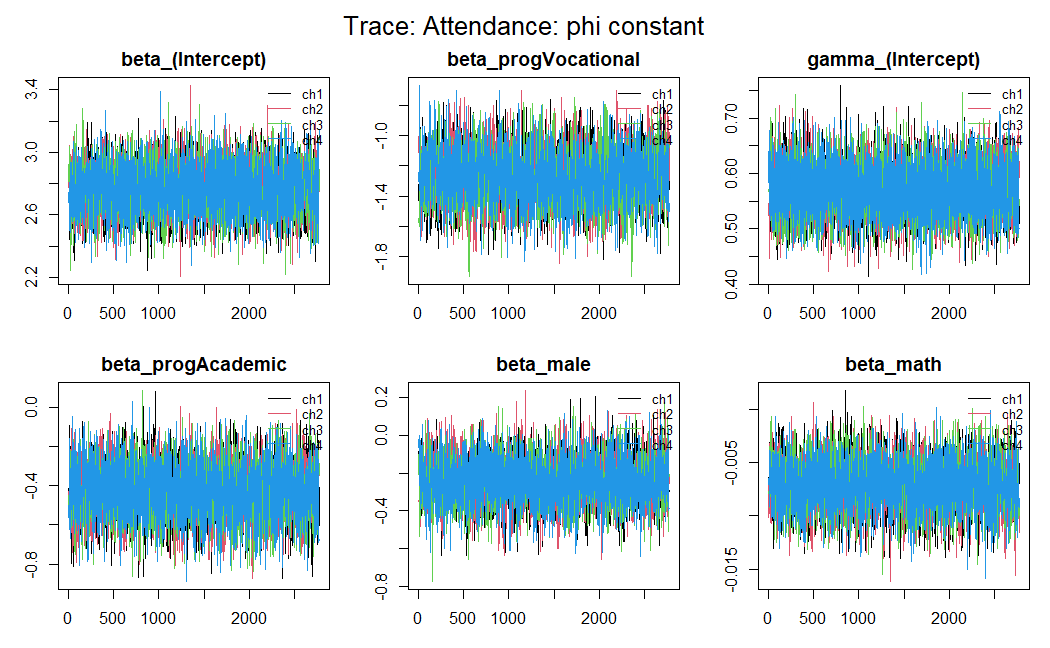}
\caption*{Trace plot}
\end{minipage}\hfill
\begin{minipage}[t]{0.58\linewidth}
\centering
\includegraphics[width=\linewidth]{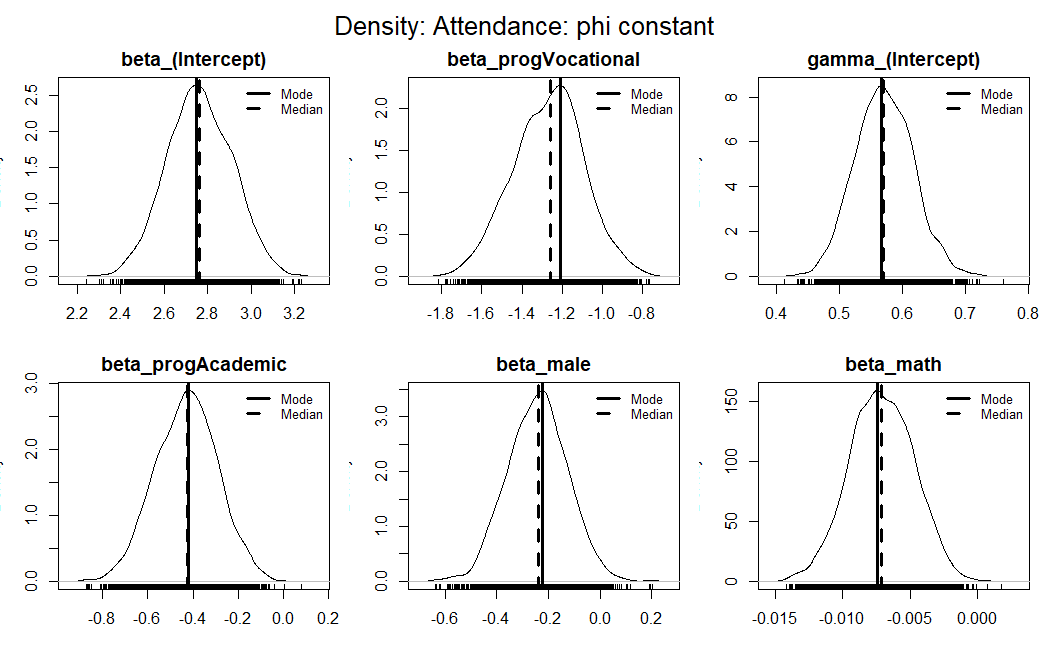}
\caption*{Density plot}
\end{minipage}

\vspace{0.8em}

\begin{minipage}[t]{0.58\linewidth}
\centering
\includegraphics[width=\linewidth]{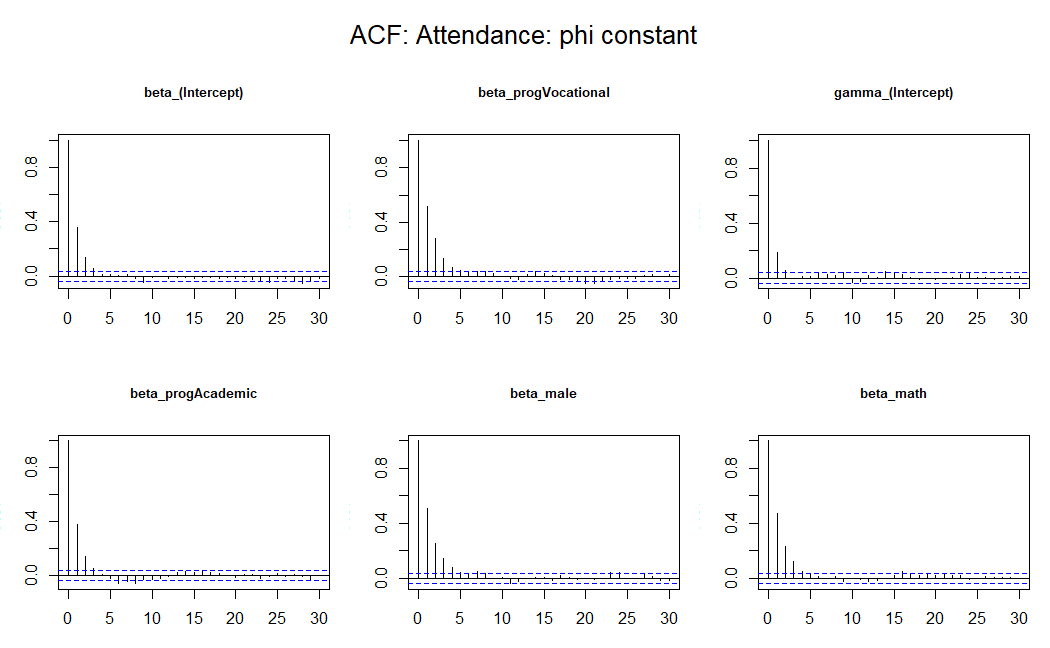}
\caption*{ACF plot}
\end{minipage}\hfill

\caption{Attendance model MCMC diagnostics under the Bayesian AETEL fit with constant dispersion.}
\label{fig:attendance_bayes_diagnostics_const}
\end{figure}

\begin{table}[H]
\centering
\caption{Attendance MCMC convergence diagnostics under constant dispersion.}
\label{tab:attendance_constant_diag}
\small
\begin{tabular}{lrrr}
\toprule
Term & ESS & $\widehat R$ & Upper CL \\
\midrule
$\beta_0$              & 4902.77 & 1.00 & 1.00 \\
$\beta_{\rm male}$     & 3575.24 & 1.00 & 1.00 \\
$\beta_{\rm Academic}$ & 4950.52 & 1.00 & 1.00 \\
$\beta_{\rm Vocational}$ & 3179.66 & 1.00 & 1.00 \\
$\beta_{\rm math}$     & 4139.00 & 1.00 & 1.00 \\
$\gamma_0$             & 6943.82 & 1.00 & 1.00 \\
\bottomrule
\end{tabular}

\par\smallskip
\begin{minipage}{0.92\textwidth}\footnotesize
\textit{Note.} Four chains were used, with 2,777 retained draws per chain
and thinning 1. The elapsed runtime was 263.08 seconds. The minimum ESS
across coefficients was 3179.66, corresponding to approximately 725.2
effective draws per minute.
\end{minipage}
\end{table}

\begin{figure}[H]
\centering

\begin{minipage}[t]{0.58\linewidth}
\centering
\includegraphics[width=\linewidth]{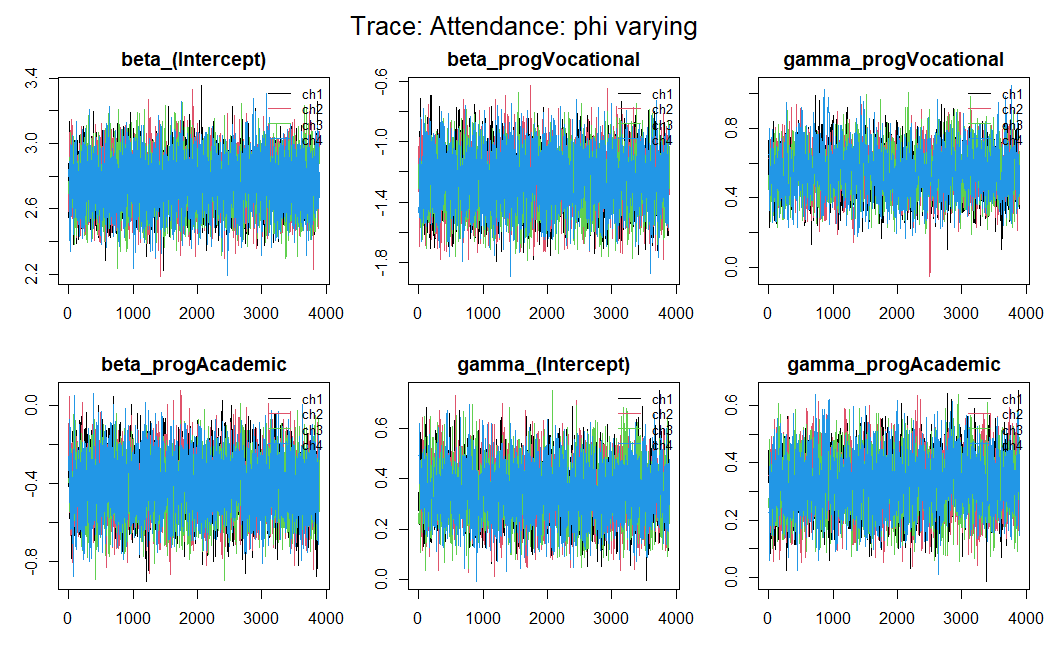}
\caption*{Trace plot}
\end{minipage}\hfill
\begin{minipage}[t]{0.58\linewidth}
\centering
\includegraphics[width=\linewidth]{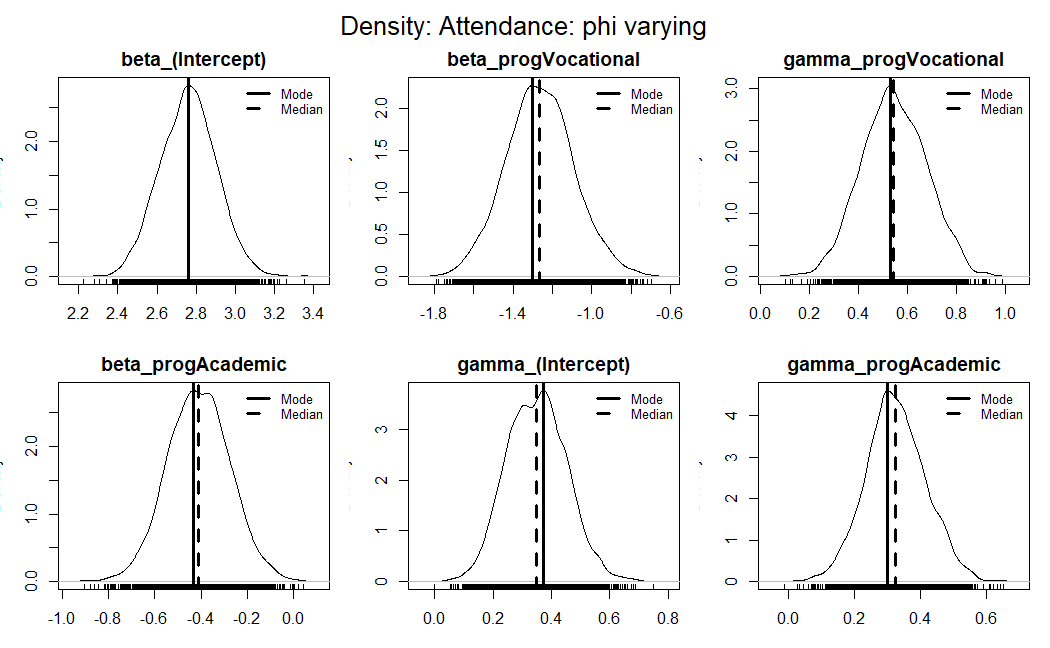}
\caption*{Density plot}
\end{minipage}

\vspace{0.8em}

\begin{minipage}[t]{0.58\linewidth}
\centering
\includegraphics[width=\linewidth]{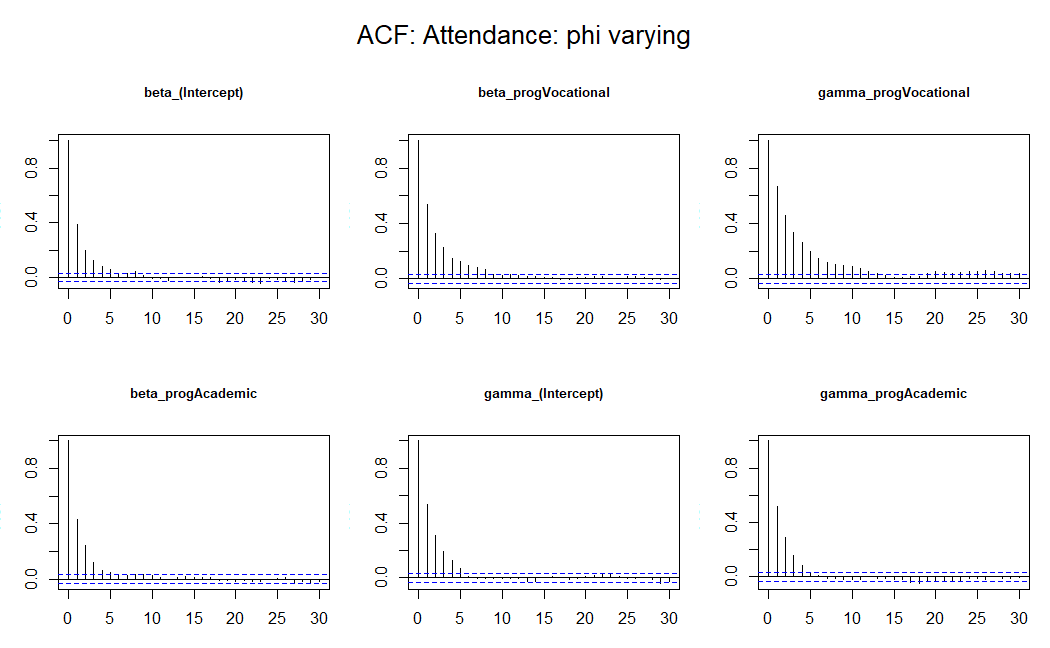}
\caption*{ACF plot}
\end{minipage}\hfill

\caption{Attendance model MCMC diagnostics under the Bayesian AETEL fit with varying dispersion.}
\label{fig:attendance_bayes_diagnostics_vary}
\end{figure}

\begin{table}[H]
\centering
\caption{Attendance MCMC convergence diagnostics under varying dispersion.}
\label{tab:attendance_vary_diag}
\small
\begin{tabular}{lrrr}
\toprule
Term & ESS & $\widehat R$ & Upper CL \\
\midrule
$\beta_0$               & 6042.89 & 1.00 & 1.00 \\
$\beta_{\rm male}$      & 4085.60 & 1.00 & 1.00 \\
$\beta_{\rm Academic}$  & 5212.91 & 1.00 & 1.00 \\
$\beta_{\rm Vocational}$ & 3766.34 & 1.00 & 1.00 \\
$\beta_{\rm math}$      & 4527.35 & 1.00 & 1.00 \\
$\gamma_0$              & 3909.57 & 1.00 & 1.00 \\
$\gamma_{\rm male}$     & 2843.77 & 1.00 & 1.01 \\
$\gamma_{\rm Academic}$ & 3925.42 & 1.00 & 1.00 \\
$\gamma_{\rm Vocational}$ & 2527.31 & 1.00 & 1.00 \\
$\gamma_{\rm math}$     & 3067.87 & 1.00 & 1.00 \\
\bottomrule
\end{tabular}

\par\smallskip
\begin{minipage}{0.92\textwidth}\footnotesize
\textit{Note.} Four chains were used, with 3,888 retained draws per chain
and thinning 1. The elapsed runtime was 664.44 seconds. The minimum ESS
across coefficients was 2527.31, corresponding to approximately 228.2
effective draws per minute.
\end{minipage}
\end{table}

Figures~\ref{fig:attendance_bayes_diagnostics_const} and
\ref{fig:attendance_bayes_diagnostics_vary} summarize the Bayesian AETEL
diagnostics for the Attendance data under the constant and varying dispersion
fits. The trace plots show stable behavior across all four chains, with no
visible drift and substantial overlap among chains. The posterior density
plots are smooth and unimodal for the displayed coefficients and the
autocorrelation functions decay fairly quickly after the first few lags.
Together with the pooled \(\widehat R\) values near one, these plots support
satisfactory convergence of the MCMC algorithm for both specifications.
The fit diagnostics based on the studentized residuals also suggest that the
model captures the main structure in the data. For both fits, the residuals
are centered close to zero and most observations lie within a moderate range,
with only a few larger positive residuals.

\setcounter{figure}{0}
\section{Takeover Bids MCMC Diagnostics}
\label{app:takeoverdiag}

\begin{figure}[H]
\centering

\begin{minipage}[t]{0.58\linewidth}
\centering
\includegraphics[width=\linewidth]{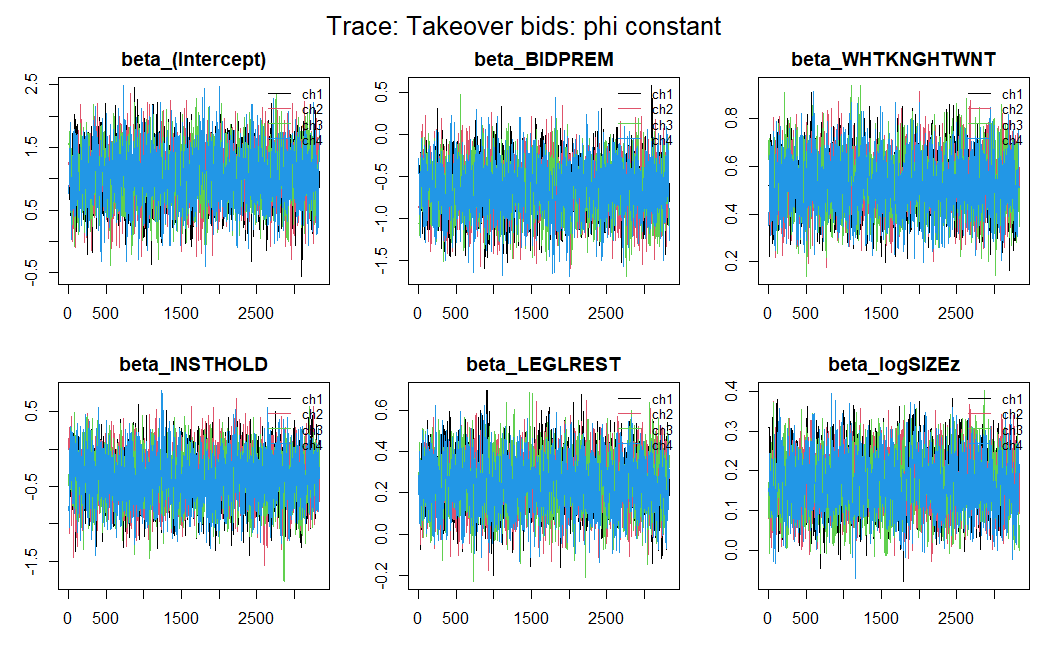}
\caption*{Trace plot}
\end{minipage}\hfill
\begin{minipage}[t]{0.58\linewidth}
\centering
\includegraphics[width=\linewidth]{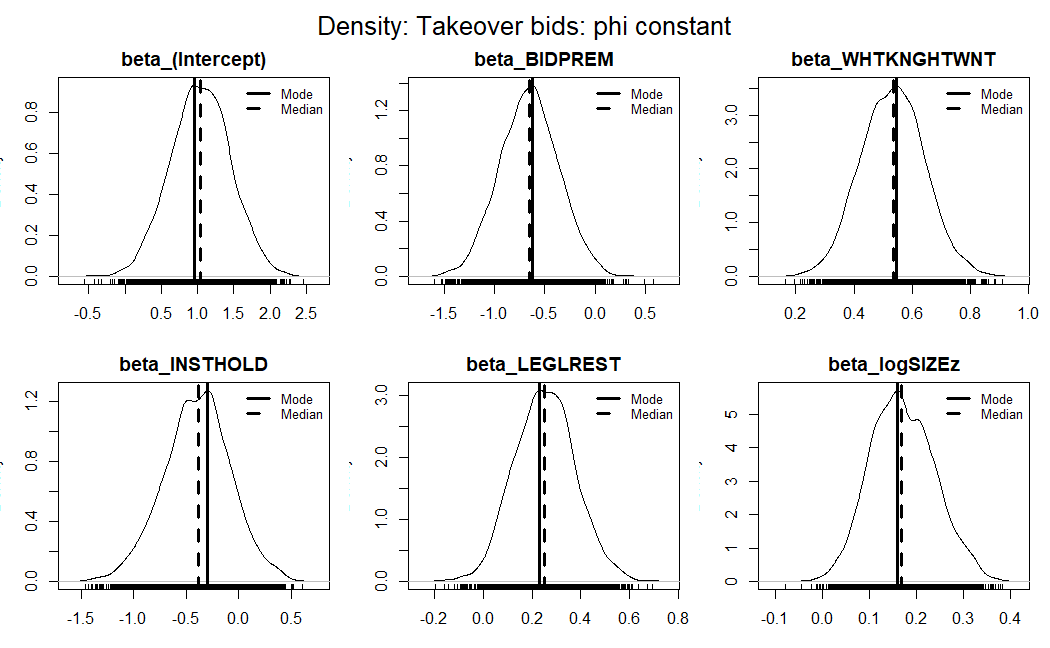}
\caption*{Density plot}
\end{minipage}

\vspace{0.8em}

\begin{minipage}[t]{0.50\linewidth}
\centering
\includegraphics[width=\linewidth]{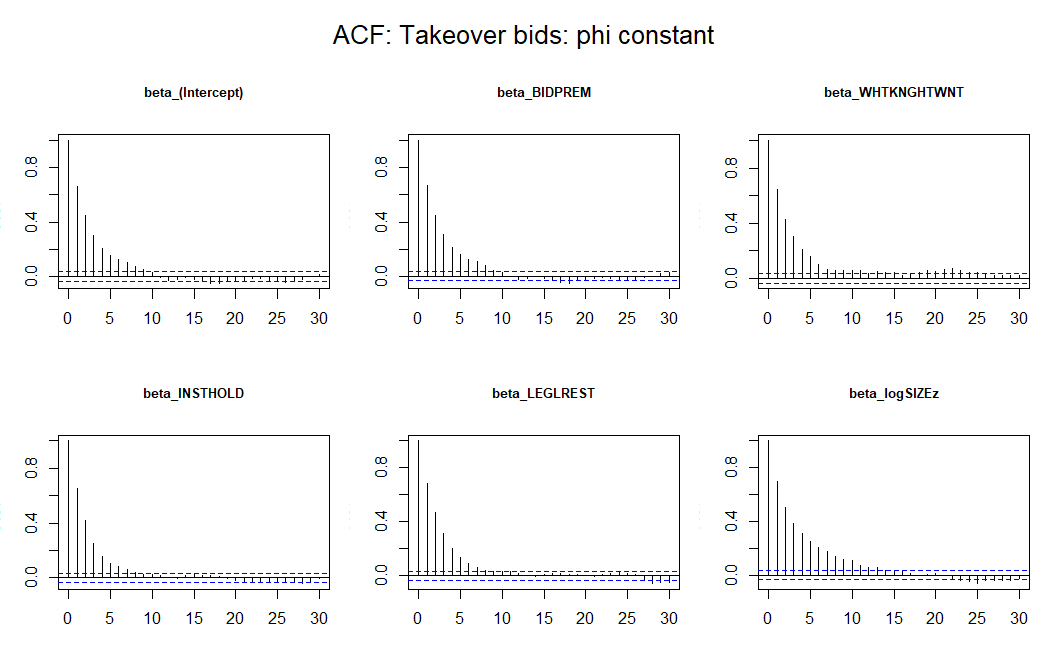}
\caption*{ACF plot}
\end{minipage}\hfill

\caption{Takeover bids model MCMC diagnostics under the Bayesian AETEL fit with constant dispersion.}
\label{fig:takeover_bayes_diagnostics_const}
\end{figure}

\begin{table}[H]
\centering
\caption{Takeover bids MCMC convergence diagnostics under constant dispersion.}
\label{tab:takeover_constant_diag}
\small
\begin{tabular}{lrrr}
\toprule
Term & ESS & $\widehat R$ & Upper CL \\
\midrule
$\beta_0$              & 2559.24 & 1.00 & 1.00 \\
$\beta_{\rm BIDPREM}$  & 2430.57 & 1.00 & 1.00 \\
$\beta_{\rm INSTHOLD}$ & 2543.93 & 1.00 & 1.00 \\
$\beta_{\rm LEGLREST}$ & 2340.55 & 1.00 & 1.00 \\
$\beta_{\rm REALREST}$ & 1637.40 & 1.00 & 1.00 \\
$\beta_{\rm FINREST}$  & 1581.64 & 1.00 & 1.00 \\
$\beta_{\rm REGULATN}$ & 2105.61 & 1.00 & 1.01 \\
$\beta_{\rm WHTKNGHT}$ & 2633.04 & 1.00 & 1.00 \\
$\beta_{\rm logSIZE}$  & 1902.26 & 1.00 & 1.00 \\
$\gamma_0$             & 1874.59 & 1.00 & 1.00 \\
\bottomrule
\end{tabular}

\par\smallskip
\begin{minipage}{0.92\textwidth}\footnotesize
\textit{Note.} We used four chains, with 3,333 retained draws per chain
and thinning 1. The elapsed runtime was 1,304.42 seconds. The minimum ESS
across coefficients was 1581.64, corresponding to approximately 72.8
effective draws per minute.
\end{minipage}
\end{table}

\begin{figure}[H]
\centering

\begin{minipage}[t]{0.58\linewidth}
\centering
\includegraphics[width=\linewidth]{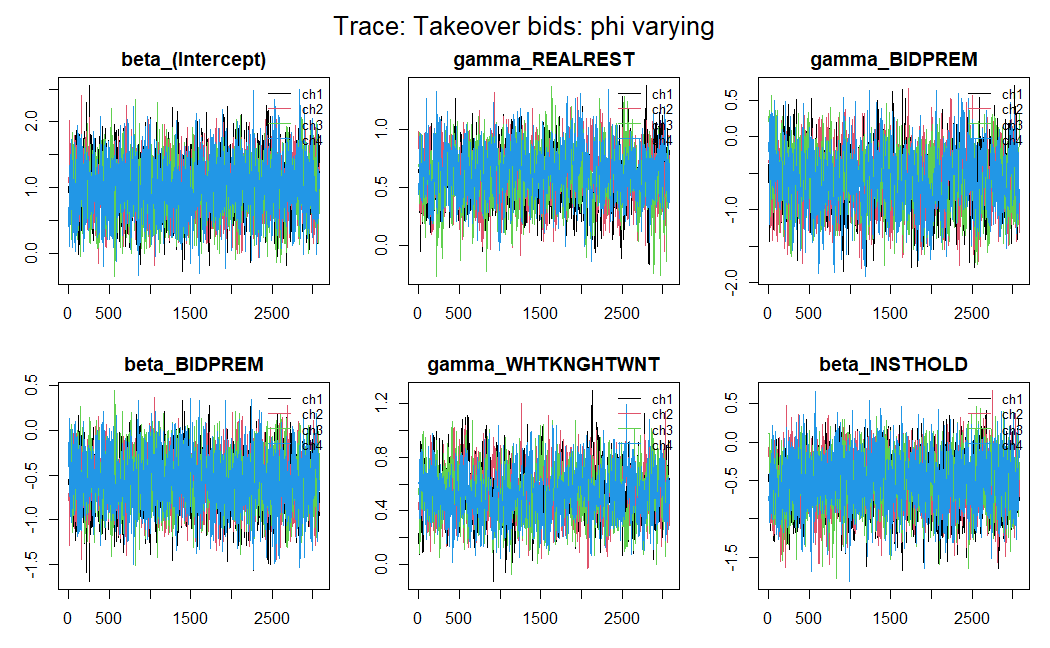}
\caption*{Trace plot}
\end{minipage}\hfill
\begin{minipage}[t]{0.58\linewidth}
\centering
\includegraphics[width=\linewidth]{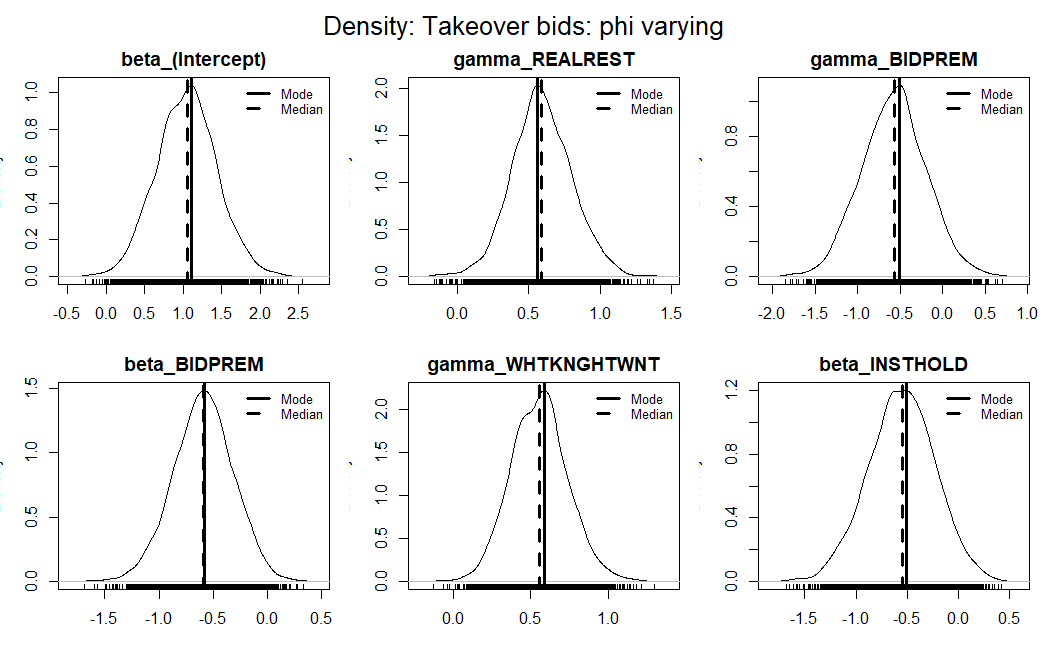}
\caption*{Density plot}
\end{minipage}

\vspace{0.8em}

\begin{minipage}[t]{0.58\linewidth}
\centering
\includegraphics[width=\linewidth]{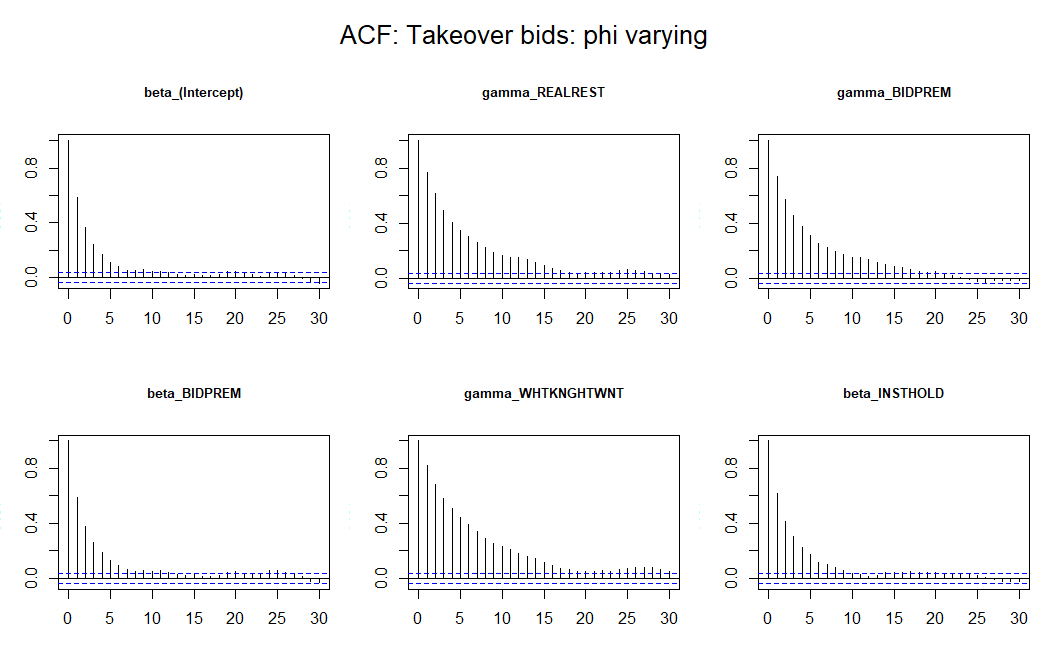}
\caption*{ACF plot}
\end{minipage}\hfill

\caption{Takeover bids model MCMC diagnostics under the Bayesian AETEL fit with varying dispersion.}
\label{fig:takeover_bayes_diagnostics_vary}
\end{figure}

\begin{table}[H]
\centering
\caption{Takeover bids MCMC convergence diagnostics under varying dispersion.}
\label{tab:takeover_vary_diag}
\small
\begin{tabular}{lrrr}
\toprule
Term & ESS & $\widehat R$ & Upper CL \\
\midrule
$\beta_0$               & 2838.81 & 1.00 & 1.01 \\
$\beta_{\rm BIDPREM}$   & 2635.56 & 1.00 & 1.01 \\
$\beta_{\rm INSTHOLD}$  & 2501.39 & 1.00 & 1.01 \\
$\beta_{\rm LEGLREST}$  & 2473.87 & 1.00 & 1.00 \\
$\beta_{\rm REALREST}$  & 1612.26 & 1.00 & 1.00 \\
$\beta_{\rm FINREST}$   & 1444.31 & 1.00 & 1.00 \\
$\beta_{\rm REGULATN}$  & 1972.07 & 1.00 & 1.00 \\
$\beta_{\rm WHTKNGHT}$  & 2196.46 & 1.00 & 1.00 \\
$\beta_{\rm logSIZE}$   & 1884.29 & 1.00 & 1.01 \\
$\gamma_0$              & 1360.92 & 1.00 & 1.00 \\
$\gamma_{\rm BIDPREM}$  & 1403.50 & 1.00 & 1.00 \\
$\gamma_{\rm INSTHOLD}$ & 1807.97 & 1.00 & 1.00 \\
$\gamma_{\rm LEGLREST}$ & 1174.43 & 1.00 & 1.00 \\
$\gamma_{\rm REALREST}$ & 1211.36 & 1.01 & 1.02 \\
$\gamma_{\rm FINREST}$  & 1979.15 & 1.00 & 1.01 \\
$\gamma_{\rm REGULATN}$ & 1233.71 & 1.00 & 1.01 \\
$\gamma_{\rm WHTKNGHT}$ & 1147.12 & 1.01 & 1.02 \\
$\gamma_{\rm logSIZE}$  & 2412.08 & 1.00 & 1.00 \\
\bottomrule
\end{tabular}

\par\smallskip
\begin{minipage}{0.92\textwidth}\footnotesize
\textit{Note.} We used four chains and with 3,333 retained draws per chain
and thinning 1. The elapsed runtime was 3,033.64 seconds. The minimum ESS
across coefficients was 1147.12, which corresponds to approximately 22.7
effective draws per minute.
\end{minipage}
\end{table}

Figures~\ref{fig:takeover_bayes_diagnostics_const} and
\ref{fig:takeover_bayes_diagnostics_vary} show stable MCMC behavior for both
dispersion specifications. The trace plots mix well across the four chains,
the posterior densities are smooth and unimodal, and the autocorrelation
functions decay in the usual way. The fit diagnostics suggest no major
systematic pattern left in the residuals, and also show why the varying
dispersion fit is useful for capturing changes in conditional spread across
firms.

\setcounter{figure}{0}
\section{Cotton Bolls MCMC Diagnostics}
\label{app:cottondiag}

\begin{figure}[H]
\centering

\begin{minipage}[t]{0.58\linewidth}
\centering
\includegraphics[width=\linewidth]{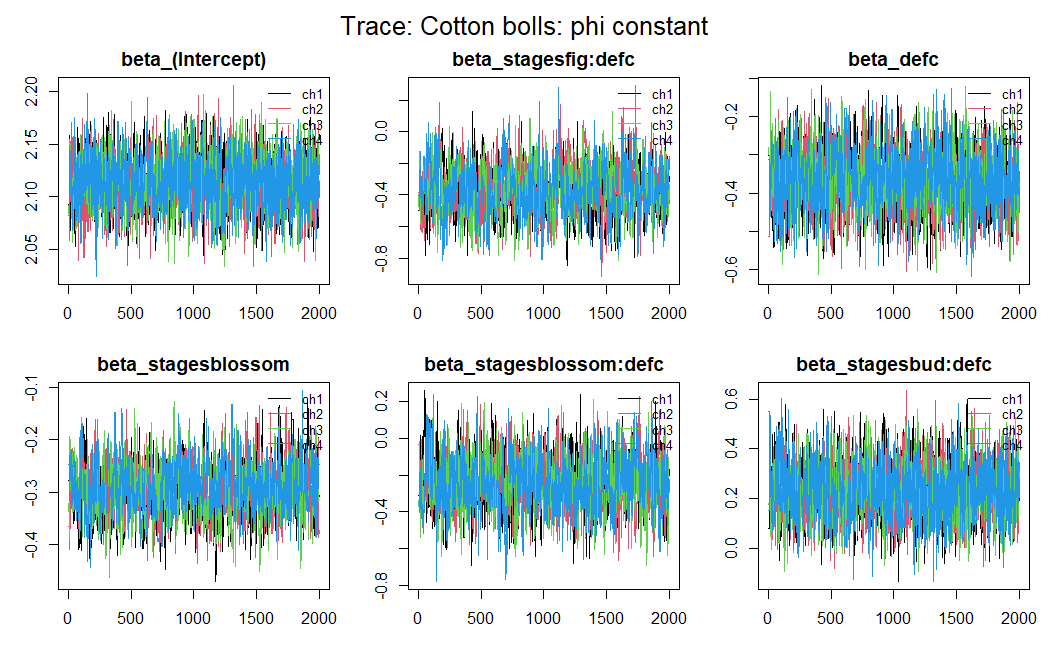}
\caption*{Trace plot}
\end{minipage}\hfill
\begin{minipage}[t]{0.50\linewidth}
\centering
\includegraphics[width=\linewidth]{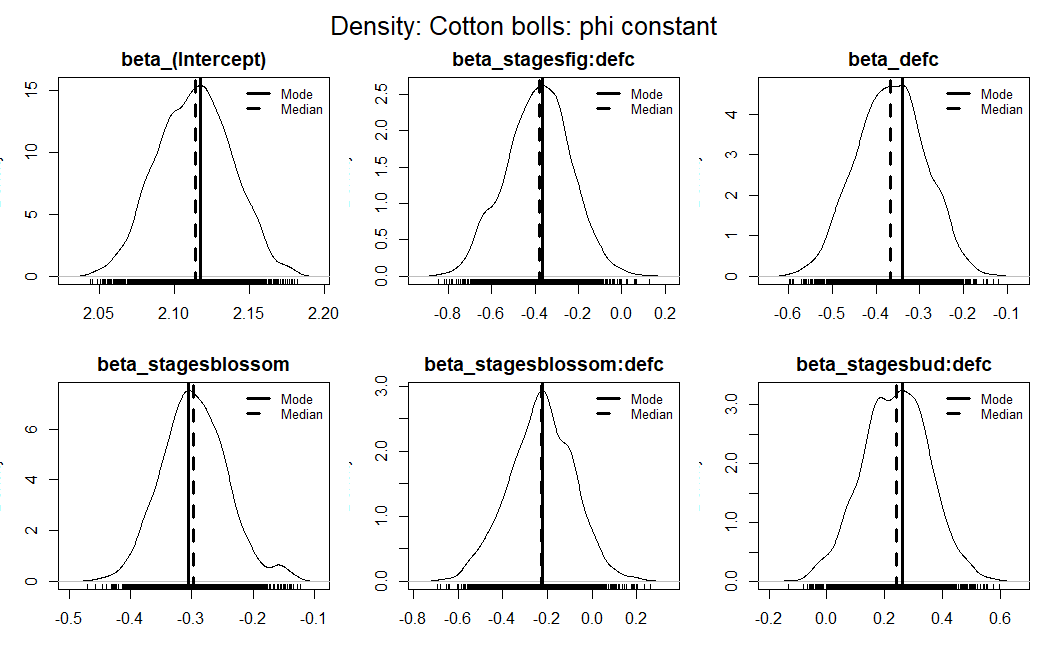}
\caption*{Density plot}
\end{minipage}

\vspace{0.8em}

\begin{minipage}[t]{0.58\linewidth}
\centering
\includegraphics[width=\linewidth]{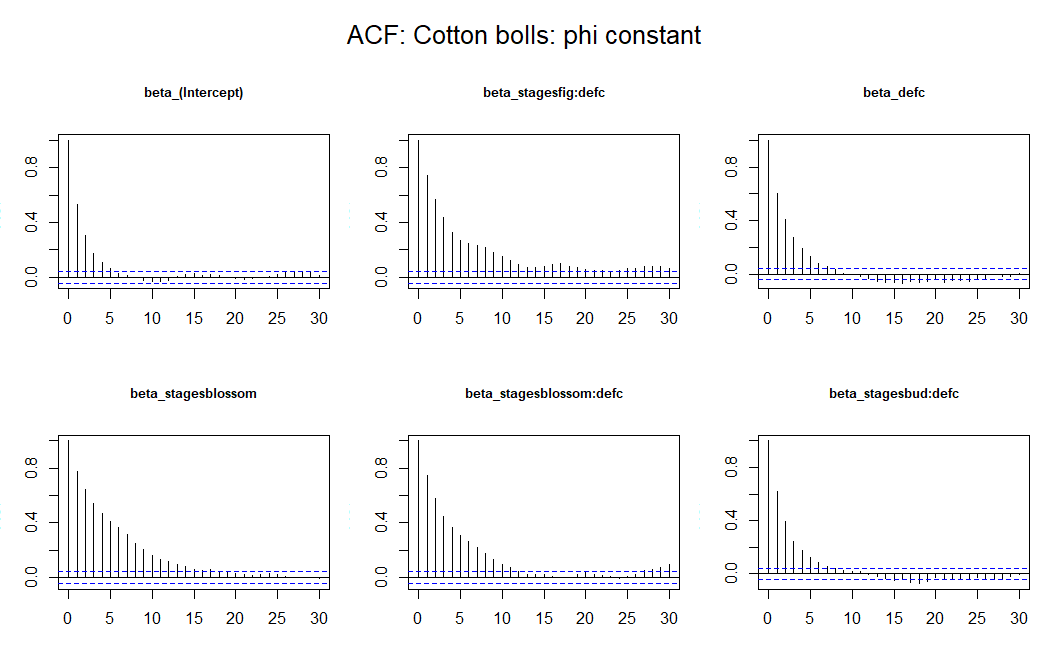}
\caption*{ACF plot}
\end{minipage}\hfill

\caption{Cotton bolls model MCMC diagnostics under the Bayesian AETEL fit with constant dispersion.}
\label{fig:cotton_bayes_diagnostics_const}
\end{figure}

\begin{table}[H]
\centering
\caption{Cotton bolls MCMC convergence diagnostics under constant dispersion.}
\label{tab:cotton_constant_diag}
\small
\begin{tabular}{lrrr}
\toprule
Term & ESS & $\widehat R$ & Upper CL \\
\midrule
$\beta_0$                    & 2068.28 & 1.00 & 1.00 \\
$\beta_{\rm bud}$            & 1790.01 & 1.00 & 1.00 \\
$\beta_{\rm blossom}$        & 885.15  & 1.00 & 1.01 \\
$\beta_{\rm fig}$            & 475.64  & 1.01 & 1.02 \\
$\beta_{\rm boll}$           & 1741.75 & 1.00 & 1.00 \\
$\beta_{\rm defc}$           & 1847.82 & 1.00 & 1.00 \\
$\beta_{\rm bud:defc}$       & 1871.26 & 1.00 & 1.00 \\
$\beta_{\rm blossom:defc}$   & 1059.44 & 1.00 & 1.01 \\
$\beta_{\rm fig:defc}$       & 761.06  & 1.01 & 1.02 \\
$\beta_{\rm boll:defc}$      & 1617.77 & 1.00 & 1.01 \\
$\gamma_0$                   & 1394.19 & 1.00 & 1.01 \\
\bottomrule
\end{tabular}

\par\smallskip
\begin{minipage}{0.92\textwidth}\footnotesize
\textit{Note.} Four chains were used, with 2,000 kept draws per chain
and thinning 1. The elapsed runtime was 644.32 seconds. The minimum ESS
across coefficients was 475.64, corresponding to approximately 44.3
effective draws per minute.
\end{minipage}
\end{table}

\begin{figure}[H]
\centering

\begin{minipage}[t]{0.55\linewidth}
\centering
\includegraphics[width=\linewidth]{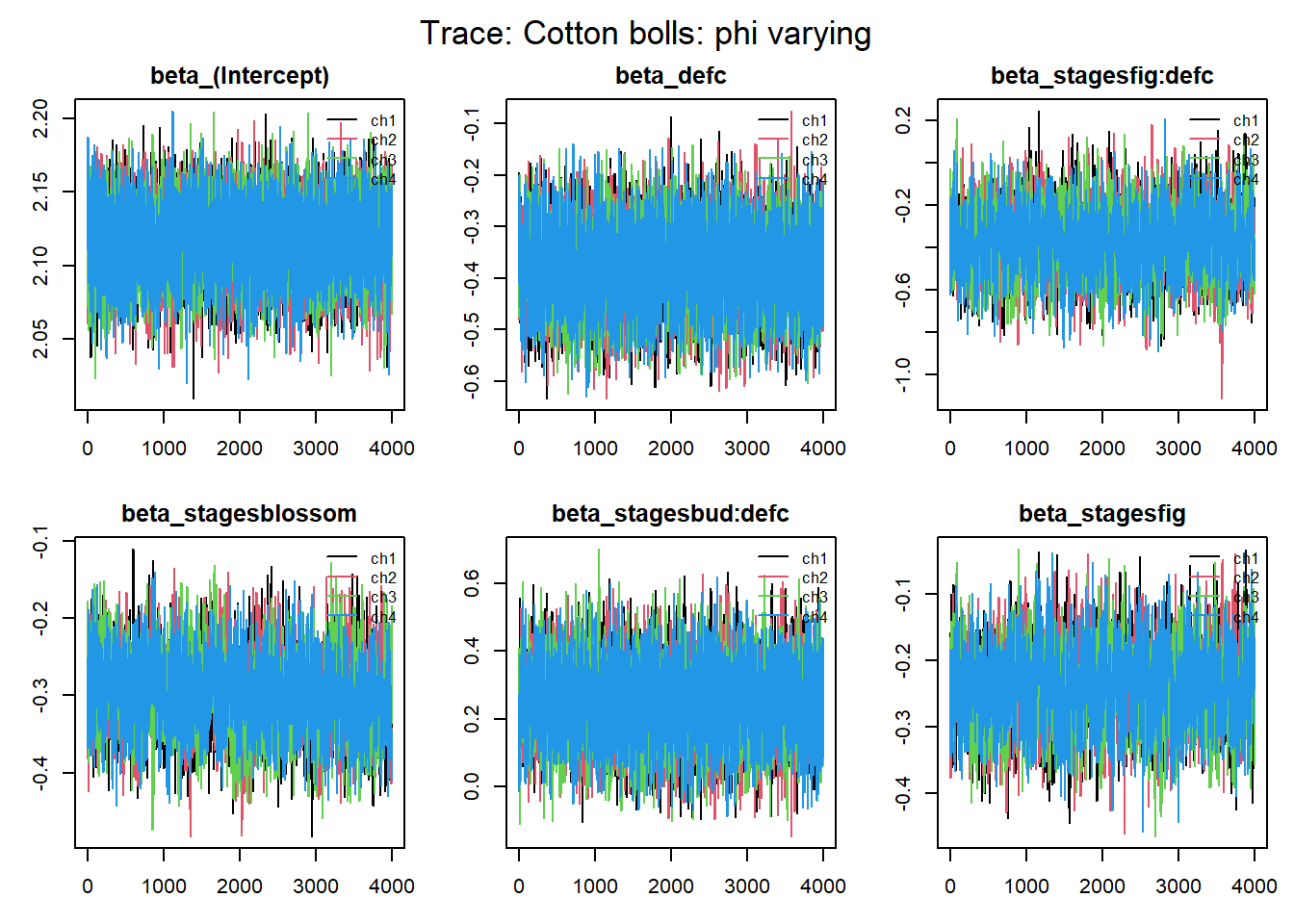}
\caption*{Trace plot}
\end{minipage}\hfill
\begin{minipage}[t]{0.55\linewidth}
\centering
\includegraphics[width=\linewidth]{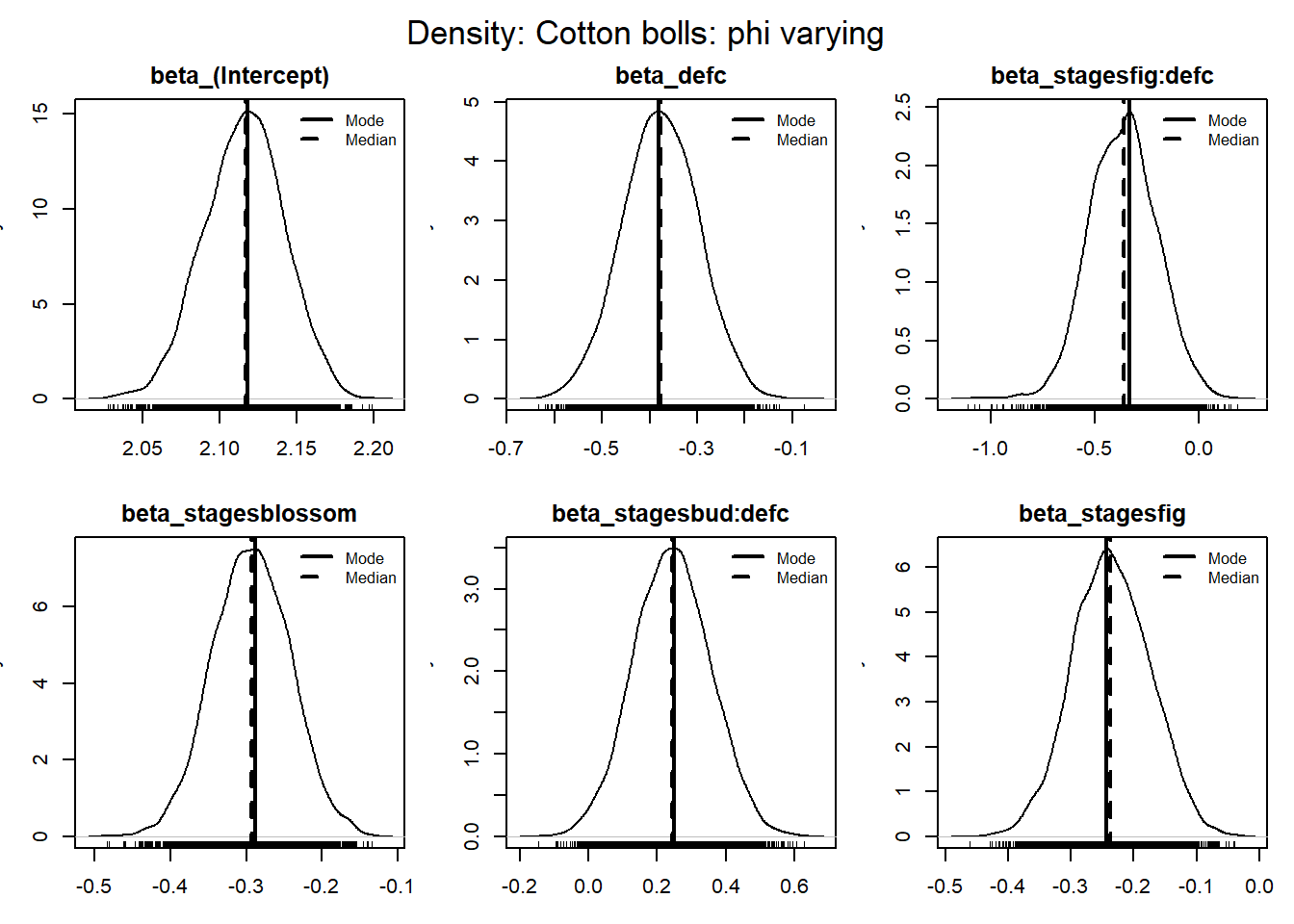}
\caption*{Density plot}
\end{minipage}

\vspace{0.8em}

\begin{minipage}[t]{0.55\linewidth}
\centering
\includegraphics[width=\linewidth]{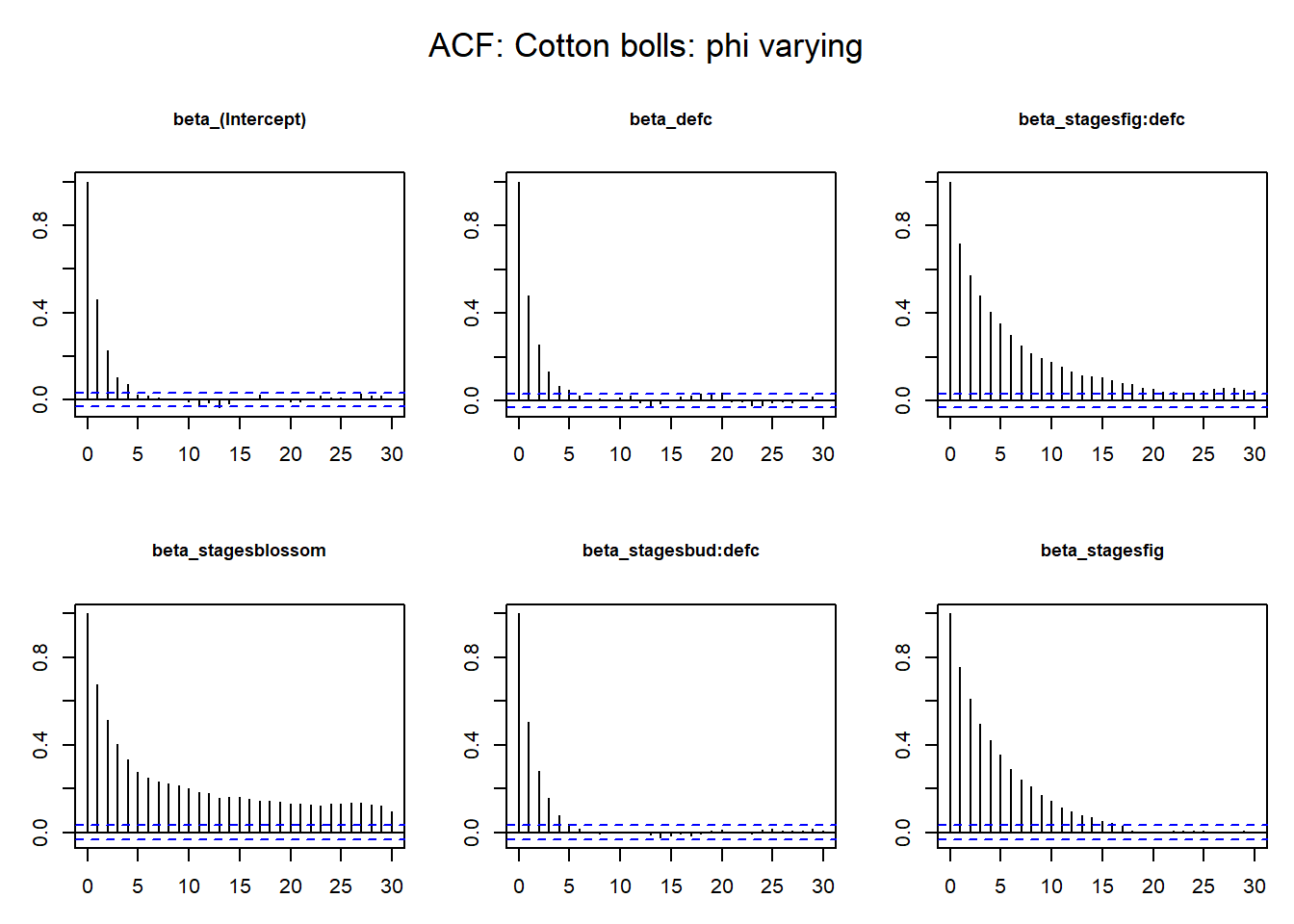}
\caption*{ACF plot}
\end{minipage}\hfill

\caption{Cotton bolls model MCMC diagnostics under the Bayesian AETEL fit with varying dispersion.}
\label{fig:cotton_bayes_diagnostics_vary}
\end{figure}

\begin{table}[H]
\centering
\caption{Cotton bolls MCMC convergence diagnostics under varying dispersion.}
\label{tab:cotton_vary_diag}
\small
\begin{tabular}{lrrr}
\toprule
Term & ESS & $\widehat R$ & Upper CL \\
\midrule
$\beta_0$                    & 5754.66 & 1.00 & 1.00 \\
$\beta_{\rm bud}$            & 4050.73 & 1.00 & 1.00 \\
$\beta_{\rm blossom}$        & 1651.61 & 1.00 & 1.00 \\
$\beta_{\rm fig}$            & 1774.14 & 1.00 & 1.01 \\
$\beta_{\rm boll}$           & 4799.99 & 1.00 & 1.00 \\
$\beta_{\rm defc}$           & 5933.45 & 1.00 & 1.00 \\
$\beta_{\rm bud:defc}$       & 5037.00 & 1.00 & 1.00 \\
$\beta_{\rm blossom:defc}$   & 978.85  & 1.00 & 1.01 \\
$\beta_{\rm fig:defc}$       & 1704.43 & 1.00 & 1.00 \\
$\beta_{\rm boll:defc}$      & 4952.97 & 1.00 & 1.00 \\
$\gamma_0$                   & 3782.47 & 1.00 & 1.00 \\
$\gamma_{\rm bud}$           & 3321.21 & 1.00 & 1.00 \\
$\gamma_{\rm blossom}$       & 546.80  & 1.00 & 1.01 \\
$\gamma_{\rm fig}$           & 316.64  & 1.00 & 1.01 \\
$\gamma_{\rm boll}$          & 3858.99 & 1.00 & 1.00 \\
$\gamma_{\rm defc}$          & 4425.79 & 1.00 & 1.00 \\
$\gamma_{\rm bud:defc}$      & 4261.66 & 1.00 & 1.00 \\
$\gamma_{\rm blossom:defc}$  & 320.08  & 1.01 & 1.02 \\
$\gamma_{\rm fig:defc}$      & 279.73  & 1.01 & 1.02 \\
$\gamma_{\rm boll:defc}$     & 4456.35 & 1.00 & 1.00 \\
\bottomrule
\end{tabular}

\par\smallskip
\begin{minipage}{0.92\textwidth}\footnotesize
\textit{Note.} Four chains were used, with 4,000 retained draws per chain
and thinning 1. The elapsed runtime was 14,407.74 seconds. The minimum ESS
across coefficients was 279.73, corresponding to approximately 1.16
effective draws per minute.
\end{minipage}
\end{table}

Figures~\ref{fig:cotton_bayes_diagnostics_const} and
\ref{fig:cotton_bayes_diagnostics_vary} summarize the Bayesian AETEL
diagnostics for the Cotton Bolls data under constant and varying dispersion.
For both models, the trace plots are stable and show no visible drift, the
posterior density plots are unimodal, and the autocorrelation functions decay
as the lag increases. This supports good mixing of the four chains and
suggests that the posterior summaries in Table~\ref{tab:cotton} are reliable.
The model fit diagnostics show a useful difference between the two
specifications. Under the constant dispersion fit, the residual cloud against
the fitted mean still shows mild structure and the scale location panel
suggests that the spread is not fully constant over the range of fitted
values. The varying dispersion fit reduces that structure and gives a more
satisfactory representation of the changing conditional variability.

\bibliographystyle{plainnat}
\bibliography{references}

\end{document}